\documentclass[final,3p,times,sort,numbers,compress]{elsarticle}

\usepackage{amssymb}

\usepackage{amsmath}
\usepackage{amsfonts}
\usepackage{braket}
\usepackage{hyperref}
\usepackage{tikz}

\usepackage{float}

\usepackage{lipsum}

\newcommand{\ketbra}[2]{\mathinner{|{#1}\rangle \hspace{-0.1pt}\langle{#2}|}}

\journal{Physics Open}

\begin{document}

\begin{frontmatter}



\title{Rabi oscillations and entanglement between two atoms interacting by the Rydberg blockade studied by the Jaynes-Cummings Model}


\author[add]{Francisco D. Santillan}
\ead{francisco.santillan01@utrgv.edu}

\author[add]{Andreas Hanke\corref{cor}}
\ead{andreas.hanke@utrgv.edu}
\cortext[cor]{Corresponding author}

\affiliation[add]{organization={Department of Physics and Astronomy, University of Texas Rio Grande Valley}, 
            addressline={1 W University Blvd},
            city={Brownsville},
            postcode={TX 78520},
            country={USA}}

\begin{abstract}
The interaction between atoms and a quantized radiation field is fundamentally important 
in quantum optics and quantum information science.  
Due to their unusual properties, Rydberg atoms are promising building blocks for two-qubit gates and 
atom-light quantum interfaces, exploiting the Rydberg blockade interaction which
prevents two atoms at close distance ($<$ 10 $\mu$m) from being simultaneously excited to Rydberg states.  
Recently, this effect was used to engineer quantum processors based on arrays of 
interacting Rydberg atoms illuminated by Raman lasers. 
Motivated by these experiments, 
we extend the Jaynes-Cummings model to study the interaction between
two Rydberg atoms interacting by the Rydberg blockade and a quantized 
radiation field. 
We consider both number (Fock) states of the field and single-mode quantum coherent states.
In particular, we discuss different types of entanglements between various 
components of the total system consisting of the two Rydberg-interacting atoms and coherent 
states of the field, and show that the behavior is significantly different compared to a system with 
non-interacting atoms corresponding to the two-atom Tavis-Cummings model. 
Our results are relevant in view of atom-light quantum interfaces
as components for future long-distance quantum communication. 
\end{abstract}




\end{frontmatter}




\section{Introduction}
\label{introduction}

The Jaynes-Cummings model (JCM) is a fundamental model in quantum optics that
describes the interaction of a 
two-level quantum system, such as atoms or other particles, with light under conditions where the quantum nature 
of photons is significant. Since its introduction in the 1960s \cite{Jaynes1963,Cummings1965} and 
experimental verification in the 1980s \cite{Rempe1987} (see also \cite{Brune1996}), the JCM 
has been used to describe a wide range of 
systems involving quantized light-matter interactions, including neutral atoms or ions in an
electromagnetic cavity, 
referred to as cavity quantum electrodynamics (cavity QED) \cite{Walther2006,Leibfried2003}, 
Josephson junctions in superconducting microwave cavities (circuit QED) \cite{Fink2008,Niemczyk2010,Burkard2020},
quantum dots \cite{Meier2004,Kasprzak2010,Basset2013,Burkard2020}, 
and others; see \cite{Larson2021} for a review.
Along with this development, the original JCM has been extended in various ways, 
including intensity-dependent or multiphoton couplings
\cite{Kochetov1987,Bashkirov2008,Sanchez2016,Rosas2021,Mandal2024},
multimode and multi-level atoms \cite{Messina2003},
many ($N>1$) atoms (Dicke model and Tavis-Cummings model) 
\cite{Kudryavtsev1993,Chen2010,Chilingaryan2013,Kirton2019},  
atoms interacting by dipole-dipole interactions 
\cite{Joshi1991,Zhang1993,Jahanbakhsh2021},
and others \cite{Larson2021}.

Among the striking properties of quantized light-matter interactions described by
the JCM are entanglement between the atoms(s) and the quantized radiation
\cite{Kudryavtsev1993,Dung1994,Tessier2003,Walther2006,Bashkirov2008,Chen2010,Li2013,Kumar2016,Li2019},
and the collapse and revival of periodic populations 
of atomic states (Rabi oscillations) for quantum coherent states of the 
field \cite{Eberly1980,Tessier2003,Larson2021}. 
These phenomena are entirely due to the quantum nature of the field 
and do not occur in a semiclassical treatment; 
indeed, the first experimental verification of field quantization according to the JCM
\cite{Rempe1987,Brune1996} was the observation of the collapse-revival 
phenomenon predicted in \cite{Eberly1980}. We note that in the 
Dicke model and Tavis-Cummings model, the atoms do not interact directly with one another
but may become entangled via their coupling to a common cavity field 
\cite{Kudryavtsev1993,Dung1994,Tessier2003}. 

Entanglement between qubits is a key resource in quantum computation \cite{Nielsen_Chuang_2010}. 
Therefore, the
ability to transform quantum information from stationary qubits encoded in, e.g., trapped atoms,
to photons (flying qubits) is a requirement for long-distance quantum communication and the 
distribution of entanglement over quantum networks
\cite{Bose1999,DiVincenzo2000,Duan2001,Kuzmich2003,vanderWal2003,Kimble2008,Li2013,Reiserer2015,Saffman2016,Li2019,Tiarks2019}. Since the JCM captures the physics of these systems it has been extensively studied in the context of quantum computation and related technologies; see \cite{Larson2021,Kumlin2023} for reviews.

\begin{figure}[t]
  \centering
  \includegraphics[width=0.64\textwidth]{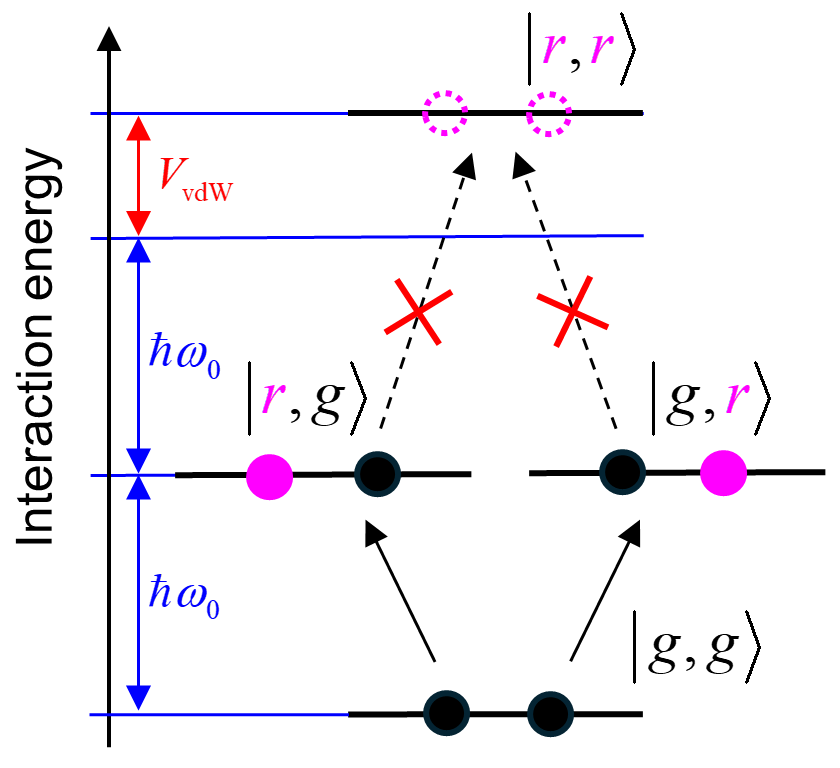}
  \caption{Rydberg blockade effect. Two atoms at close distance ($<$ 10 $\mu$m) in Rydberg states $\ket{r}$ 
  (dashed cyan circles)
  interact by van der Waals forces $V_{\text{vdW}}$, resulting in a shift of energy levels as indicated
  (shown is a case for which $V_{\text{vdW}}$ is positive, which applies to $nS_{1/2}$ Rydberg states;
  for other Rydberg states $V_{\text{vdW}}$ can be negative \cite{Reinhard2007}). As a result, only one 
  of the two atoms can be excited from the ground state $\ket{g}$ (filled black dots) to a Rydberg state 
  $\ket{r}$ (filled cyan dots) by radiation resonant to the frequency 
  $\omega_0$ of the transition $\ket{g} \leftrightarrow \ket{r}$ of an unperturbed atom, but 
  not both atoms simultaneously, if $V_{\text{vdW}}$ is larger than the width of the resonance line. 
  In this case, the doubly excited state $\ket{r,r}$ is inaccessible to the system.} 
  \label{fig:rb}
\end{figure}

\begin{figure}[t]
  \centering
  \includegraphics[width=0.58\textwidth]{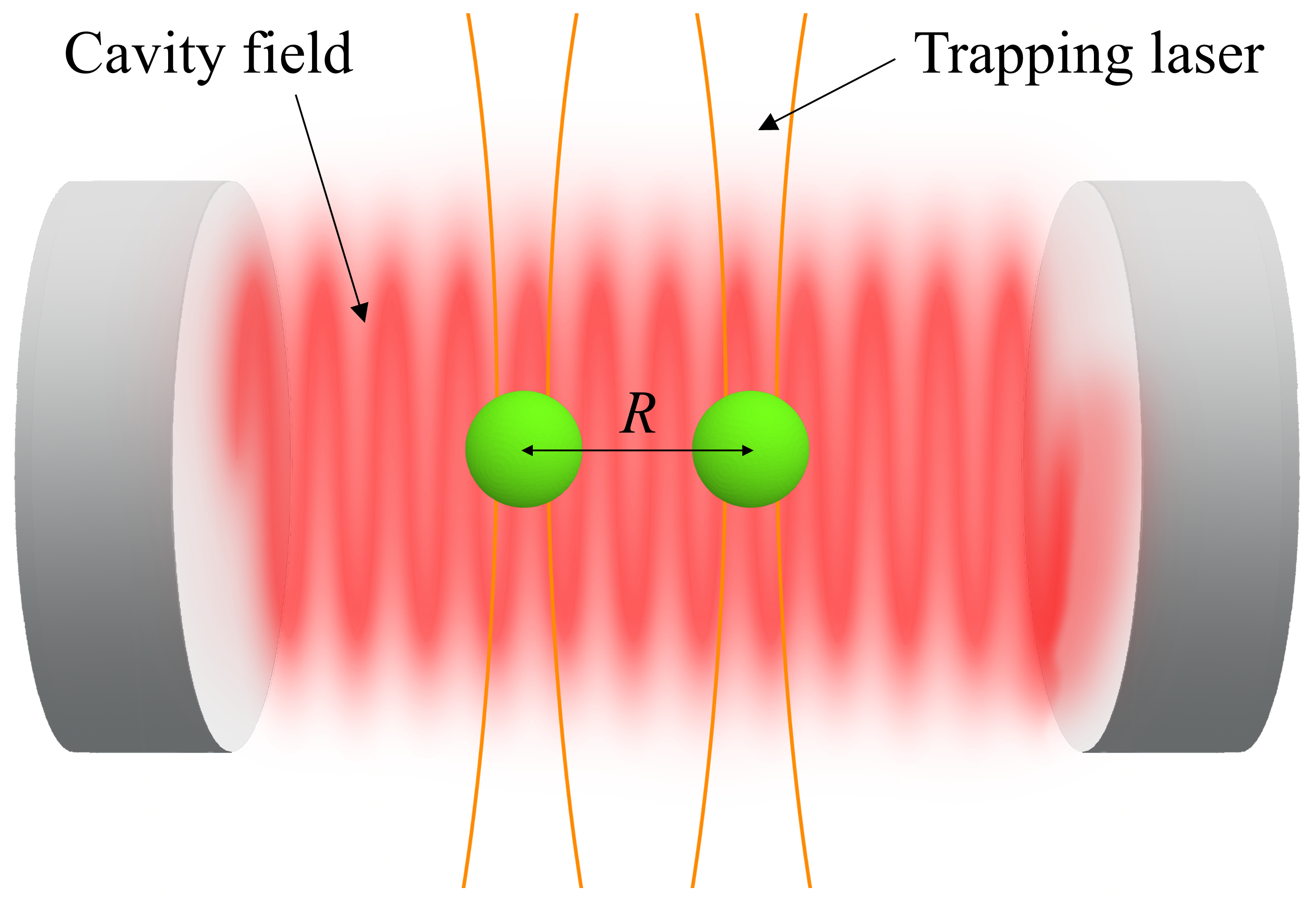}
  \caption{Schematics of a possible experimental setup for probing two atoms interacting by the Rydberg blockade and with a quantized radiation field in a cavity.
  The atoms (green balls) are trapped at close distance ($R < 10\,\mu$m) by optical tweezers in an 
  electromagnetic cavity and exposed to a radiation field resonant to the atomic transition 
  $\ket{g} \leftrightarrow \ket{r}$ to excite them to Rydberg states (Figure \ref{fig:rb}). 
  The laser beams of the optical tweezers are far detuned from the atomic transition.}
  \label{fig:atoms}
\end{figure}

An almost ideal physical setup for realizing JCM physics are Rydberg atoms interacting with a single mode of 
a cavity field 
\cite{Goy1983,Brune1996,Raimond2001,Li2013,Lee2017}. Rydberg atoms - atoms
(typically alkali) excited to states with large principal quantum numbers $n > 20$ (Rydberg states) - interact strongly with one another by van der Waals (vdW) interactions
($V_{\text{vdW}}(R) = C_6/R^6$ with $C_6 \sim n^{11}$ where $R$ is the interatomic distance) 
and with external electric fields (due to their large atomic polarizability $\alpha \sim n^7$) \cite{Gallagher1994,Reinhard2007,Sibalic2018}. 
Van der Waals and dipole-dipole energy shifts of pairs of interacting rubidium Rydberg atoms 
for different quantum numbers $n$, $\ell$, $j$, $m_j$ were obtained in \cite{Reinhard2007}.
A distinct phenomenon in ensembles of $N \ge 2$ atoms at close distance ($<$ 10 $\mu$m) is the Rydberg blockade effect \cite{Jaksch2000,Lukin2001,Saffman2005,Petrosyan2008}: 
Two atoms in a Rydberg state $\ket{r}$ interact via van der Waals forces
over distances of several $\mu$m, whereas the interaction between atoms in the ground state $\ket{g}$ is negligible; 
as a result, only one single atom within the blockade radius can be optically 
excited to a Rydberg state $\ket{r}$ by radiation resonant to the transition $\ket{g} \leftrightarrow \ket{r}$ of an unperturbed atom, because the van der Waals interaction shifts the energy levels of all other atoms within the blockade radius out of resonance (illustrated in Figure \ref{fig:rb} for two atoms). 
The Rydberg blockade effect was proposed theoretically as a way to perform fast quantum gates in 
\cite{Jaksch2000,Lukin2001,Saffman2005,Petrosyan2008} and experimentally demonstrated for two atoms individually 
trapped in optical tweezers at a distance of a few $\mu$m in \cite{Urban2009,Gaetan2009,Wilk2010}.
A CNOT gate based on the Rydberg blockade effect was experimentally demonstrated in \cite{Isenhower2010}.
In these experiments, the laser pulses used to excite the atoms 
to Rydberg states are described as classical electromagnetic waves with 
precisely controlled duration and power.
Conversely, in this paper we consider the case that the radiation is 
quantized.
Figure \ref{fig:atoms} shows a possible experimental setup for probing
two atoms interacting by the Rydberg blockade and with a quantized radiation field
in a cavity
\footnote{More generally, the quantized radiation interacting with the
atoms can be a cavity field as shown in Figure \ref{fig:atoms} or, e.g.,
propagating laser light in a quantum coherent state (Section \ref{sec:coherent}).}.

In an ensemble of $N>1$ atoms within the blockade radius, a light mode resonant to the transition 
$\ket{g} \leftrightarrow \ket{r}$ of an unperturbed atom couples the $N$-body ground state 
\begin{equation} \label{GN}
    \ket{G} = \ket{g_1, \ldots, g_N}
\end{equation}
to an entangled $N$-body state 
\begin{equation} \label{RN}
\ket{R} = \frac{1}{\sqrt{N}} \sum\limits_{j=1}^N 
e^{i {\bf k} \cdot {\bf x}_j} 
\ket{g_1, \ldots, g_{j-1}, r_j, g_{j+1}, \ldots, g_N}
\end{equation}
in which one Rydberg excitation is coherently shared between all atoms in the ensemble 
\cite{Guerlin2010,Dudin2012,Paris-Mandoki2017,Spong2021}. 
In \eqref{RN} $\ket{g_j}$ and $\ket{r_j}$ denote the ground state and a Rydberg state of atom $j$, 
respectively, ${\bf x}_j$ is the position of atom $j$, and $\bf{k}$ is the wave vector of the
light mode. 
Under conditions of a perfect Rydberg blockade any state with two or more atoms excited to $\ket{r}$ is inaccessible to the system; in this case, the ensemble of $N$ atoms is considered to be
an effective two-level Rydberg ``superatom'' with ground state $\ket{G}$ and excited state $\ket{R}$. 
However, of course, states other than $\ket{G}$ and $\ket{R}$ are possible, depending 
on the initial conditions of the system; for example, the states $\ket{r,g}$ and $\ket{g,r}$ 
can occur in the case of two atoms (Section \ref{sec:caseC}).
The collective nature of 
the excitation in $\ket{R}$ in \eqref{RN} enhances the coupling of light to a Rydberg superatom 
by a factor of $\sqrt{N}$; as a result, Rydberg systems with $N \gg 1$ (several hundreds to thousands) atoms
have attracted tremendous interest in atomic physics and quantum optics 
\cite{Guerlin2010,Dudin2012,Paris-Mandoki2017,Tiarks2019,Spong2021}.
The Rydberg blockade effect combined with the strong coupling between Rydberg superatoms and light have 
led to numerous applications of Rydberg atoms in quantum optics and quantum information science, reviewed in \cite{Saffman2016,Shi2022,Kumlin2023}.

The case of $N=2$ Rydberg atoms is of special interest because pairs of
interacting Rydberg atoms are building blocks of quantum computers in which rubidium (Rb)
atoms are trapped and transported 
by optical tweezers in reconfigurable arrays 
\cite{Levine2019,Bluvstein2021,Bluvstein2022,Evered2023,Bluvstein2024}. 
In these systems, hyperfine ground states of $^{87}$Rb 
atoms serve as qubit states $\ket{0}$, $\ket{1}$, and excitation to Rydberg states 
$\ket{r}$ by laser pulses described as classical electromagnetic waves
with precisely controlled duration and power 
is used to entangle pairs of atoms at close distances (3-5$\,\mu$m) 
by the Rydberg blockade mechanism to generate two-qubit operations. 
This process entangles qubits stored in individual atoms because atoms in the qubit state $\ket{1}$ 
are excited to the Rydberg state $\ket{r}$ (via an intermediate excited state $\ket{e}$) when illuminated 
by laser pulses of appropriate frequency (Rydberg pulses) whereas atoms in $\ket{0}$ are not excited. 
This mechanism can also entangle three or more atoms enabling native multi-qubit gates
\cite{Evered2023,Bluvstein2024}.

Motivated by these experiments, in this paper we extend the Jaynes-Cummings model to 
study two interacting Rydberg atoms under perfect Rydberg blockade conditions, so that the doubly 
excited state $\ket{r,r}$ is not accessible (Figure \ref{fig:rb}). 
This direct interaction between the atoms makes the model fundamentally different from 
the Dicke model and Tavis-Cummings model, in which the atoms do not interact directly 
with one another \cite{Kudryavtsev1993,Dung1994,Tessier2003}.
Our motivation is to
better understand the interplay between interacting Rydberg atoms and a quantized radiation field
in view of the physical implementation of quantum state-transfer between atoms and
photons for future quantum information processing
\cite{Bose1999,Duan2001,Kuzmich2003,Li2013,Kumar2016,Li2019} and photon-photon quantum gates 
\cite{Gorshkov2011,Paredes-Barato2014,Khazali2015,Tiarks2019}. 
%
%
In Section \ref{sec:hamiltonian} 
we formulate the Hamiltonian of the model and discuss its eigenstates and eigenvalues. 
We then discuss the dynamics of the system for 
different initial conditions for number (Fock) states 
of the radiation field (Section \ref{sec:number}) and a quantum coherent field (Section \ref{sec:coherent}). 
In Section \ref{sec:entanglement}, we discuss different types of entanglements between various 
components of the system. 
In Section \ref{sec:conclusions}, we summarize our results and compare the 
entanglements for a system with Rydberg-interacting atoms with a system
of non-interacting atoms corresponding to the Tavis-Cummings model. 
Detailed derivations of some results discussed
in the main text are left to \ref{sec:appA} - \ref{sec:appTC}. 


\section{Hamiltonian and eigenstates}
\label{sec:hamiltonian}

In this section, we extend the Jaynes-Cummings model for a single atom 
\footnote{The Jaynes-Cummings model for a single atom is discussed in many textbooks on quantum optics;
see, e.g., \cite{Meystre2021,Lambropoulos2007}, and the recent review in \cite{Larson2021} 
for a comprehensive list of references.}
to two Rydberg atoms interacting by the Rydberg blockade. 
We consider two two-level atoms $a$ and $b$ with ground states $\ket{g_a}$, $\ket{g_b}$
and Rydberg states $\ket{r_a}$, $\ket{r_b}$
\footnote{When there is no danger of confusion we omit the labels $a$
and $b$ and denote the ground and Rydberg states for both atoms by $\ket{g}$ and $\ket{r}$,
respectively.},
respectively, interacting  with a single mode of a quantized radiation field ($f$) with 
(angular) frequency $\omega_f$
\footnote{If not stated otherwise, we use the term ``frequency'' for the angular frequency in units of 
$\text{rad}\cdot\text{s}^{-1}$.}.
We assume that the unperturbed transition frequencies $\omega_0 = \omega_r - \omega_g$ 
of the atomic transition $\ket{g} \leftrightarrow \ket{r}$ of both atoms are equal.
The combined system of the two atoms and the 
quantized field is spanned by product states of the form
\begin{equation} \label{prod3}
    \ket{\psi} = \ket{\text{atom } a} \otimes \ket{\text{atom } b} \otimes
    \ket{\text{field}}
     = \ket{\psi_a,\psi_b,\phi_f} \, . 
\end{equation}
We assume that the Rydberg blockade between the atoms is in full effect, so that the doubly 
excited state, $\ket{r, r, \phi_f}$, is not accessible to the system. 
Thus, accessible states are spanned by states of the form 
$\ket{g,g,n}$, $\ket{g,r,n}$, and $\ket{r,g,n}$, with the photon 
number (Fock) states $\ket{n}$, $n = 0,1,2, \ldots$, of the field.
We describe the interaction between the atoms and the quantized field by a 
Jaynes-Cummings-type Hamiltonian in the rotating wave approximation,
\footnote{To simplify notation, in \eqref{ham} we use the $\otimes$ symbol only to separate operators 
acting on atoms $a$ and $b$, but not for the field operators 
$\hat{a}$, $\hat{a}^{\dag}$.} 
\begin{equation} \label{ham}
    \hat H = \hbar \omega_f {\hat a}^{\dag} {\hat a} - \frac{1}{2} \hbar \omega_0
    \left( {\hat \sigma}_z^a + {\hat \sigma}_z^b \right)
    \, + \hbar \lambda_a \left( {\hat \sigma}_{+}^a {\hat a} 
    + {\hat \sigma}_{-}^a {\hat a}^{\dag} \right) \otimes {\hat P}_g^{\,b}
    \, + {\hat P}_g^{\,a} \otimes \hbar \lambda_b \left( {\hat \sigma}_{+}^b {\hat a} 
    + {\hat \sigma}_{-}^b {\hat a}^{\dag} \right) \, .
\end{equation}
The first term in \eqref{ham} is the energy of the field with the photon number operator 
${\hat n} = {\hat a}^\dag {\hat a}$, where ${\hat a}^\dag$, $\hat a$ are bosonic creation and 
annihilation operators and $\hbar = h / \left( 2\pi \right)$
is the reduced Planck constant.
The second term in \eqref{ham} is the energy of the unperturbed atoms $a$, $b$ (upper labels)
expressed in terms of the Pauli operators 
${\hat \sigma}_z = \ketbra{g}{g} - \ketbra{r}{r}$, 
where the zero of energy is chosen as the average of the energies of $\ket{g}$ and 
$\ket{r}$.
The third and fourth terms in \eqref{ham} 
describe processes in which either atom $a$ or atom $b$ 
is excited from $\ket{g}$ to $\ket{r}$ by absorption of one photon, or decays from $\ket{r}$ to $\ket{g}$
by emission of one photon, with coupling constants $\lambda_a$, $\lambda_b$ for atoms $a$, $b$,
respectively; ${\hat \sigma}_{+} = \ketbra{r}{g}$ and 
${\hat \sigma}_{-} = \ketbra{g}{r}$ are raising and lowering Pauli operators for atoms
$a$, $b$ (upper labels in \eqref{ham}).
In the basis $\left\{ {\ket{g} , \ket{r} } \right\}$ the Pauli operators are represented by
$2 \times 2$ matrices 
\footnote{Note that in the basis $\left\{ {\ket{g} , \ket{r} } \right\}$ (in this order) 
the raising and lowering Pauli matrices $\sigma_{+}$, $\sigma_{-}$ in \eqref{pauli} are 
defined the opposite way compared to raising and lowering Pauli matrices for 
spin-$1/2$ particles.}
\begin{equation} \label{pauli}
	\sigma_z = \begin{pmatrix} 1 & 0 \\ 0 & {-1} \end{pmatrix} , \, 
    \sigma_{+} = \begin{pmatrix} 0 & 0 \\ 1 & 0 \end{pmatrix} , \,  
    \sigma_{-} = \begin{pmatrix} 0 & 1 \\ 0 & 0 \end{pmatrix} \,  .
\end{equation}
Since the doubly excited state $\ket{r,r,\phi_f}$ is not accessible to the system
due to the Rydberg blockade, excitation and decay of one atom can only occur when 
the other atom is in its ground state $\ket{g}$; this effect is incorporated in the 
Hamiltonian \eqref{ham} by the projection operators
${\hat P}_g^{\,a} = \ketbra{g_a}{g_a}$ and
${\hat P}_g^{\,b} = \ketbra{g_b}{g_b}$ to the ground states
of atoms $a$ and $b$, respectively.
The fact that the doubly excited state $\ket{r,r,\phi_f}$ is not accessible to the system
creates a direct interaction between the atoms in the Hamiltonian \eqref{ham},
in addition to the interaction of the atoms with the quantized radiation field.
This makes our model fundamentally different from previous studies using the  
Dicke model and Tavis-Cummings model, in which the projection operators
${\hat P}_g^{\,a}$ and ${\hat P}_g^{\,b}$ are absent and hence the atoms do not interact 
directly with one another \cite{Kudryavtsev1993,Dung1994,Tessier2003,Li2013}
(see Section \ref{sec:conclusions}).

Following \cite{Kudryavtsev1993} for the case of a stationary field in a cavity,
we assume that atom $a$ is located at a peak 
of the cavity field mode and that the coupling constant $\lambda_a$ in \eqref{ham}
is constant and real, i.e.,
\begin{equation} \label{la}
    \lambda_a \equiv \lambda \in \mathbb{R} \, .
\end{equation}
The coupling constant $\lambda_b$ for atom $b$ 
then has the form \cite{Kudryavtsev1993}
\begin{equation} \label{lb}
    \lambda_b = \lambda \cos(k R) \, ,
\end{equation}
where $k$ is the wave number of the cavity field mode
and $R$ is the distance between the atoms (Figure \ref{fig:atoms}). 
In this paper we assume for simplicity that $k R = \ell\,2 \pi$ with 
some integer $\ell$, so that 
\begin{equation} \label{lambda}
    \lambda_a = \lambda_b = \lambda \in \mathbb{R} \, .
\end{equation}
Since $\hat{N} = {\hat a}^{\dag} {\hat a} - \frac{1}{2}
({\hat \sigma}_z^a + {\hat \sigma}_z^b)$ 
is a constant of motion of the system
\footnote{For $\omega_f = \omega_0$ the quantity $N+1$ corresponds to the total number
of energy quanta $\hbar \omega_f$ in the system, where $N$ is an eigenvalue of $\hat{N}$.},
i.e., 
$[\hat{H},\hat{N}] = 0$, it is convenient to write $\hat{H}$ in the form
(using \eqref{lambda})
\begin{equation} \label{ham2}
    \hat H = \hbar \omega_f \hat{N}
    + \frac{1}{2} \hbar \Delta \left( {\hat \sigma}_z^a + {\hat \sigma}_z^b \right)
    \, + \hbar \lambda \Big[ \Big( {\hat \sigma}_{+}^a {\hat a} 
    + {\hat \sigma}_{-}^a {\hat a}^{\dag} \Big) \otimes {\hat P}_g^{\,b}
    \, + {\hat P}_g^{\,a} \otimes \Big( {\hat \sigma}_{+}^b {\hat a} 
    + {\hat \sigma}_{-}^b {\hat a}^{\dag} \Big) \Big] \, ,
\end{equation}
where 
\begin{equation} \label{Delta}
    \Delta = \omega_f - \omega_0
\end{equation}
is the detuning of the frequency $\omega_f$ of the field from the 
atomic resonance frequency $\omega_0$. 

Since $\hat{N}$ is a constant of motion, 
for given photon number $n=0,1,2, \ldots$, the interaction 
between the atoms and the field 
in the Hamiltonian \eqref{ham2} couples only the states
\begin{equation} \label{states3}
 \ket{\psi_1^{(n)}} := \ket{r,g,n} \, , \, \,  
 \ket{\psi_2^{(n)}} := \ket{g,r,n} \, , \, \,
 \ket{\psi_3^{(n)}} := \ket{g,g,n+1} \, .
\end{equation}
Thus, the total Hilbert space $\mathbb{H}$ of the system decouples into 
3-dimensional subspaces
$\mathbb{H}_n = \text{span} \{ \ket{\psi_1^{(n)}}, 
\ket{\psi_2^{(n)}}, \ket{\psi_3^{(n)}} \}$,
and $\hat{H}$ in \eqref{ham2} can be expressed as a direct sum 
${\hat H} = \mathop \oplus\limits_{n = 0}^\infty {\hat H}^{(n)}$
where the action of each component ${\hat H}^{(n)}$ is confined to the subspace $\mathbb{H}_n$
\footnote{Without a direct interaction between the atoms as in the 
Tavis-Cummings model, the doubly excited state $\ket{r,r,\phi_f}$ is accessible to the system.
As a result, the system decouples into 4-dimensional subspaces 
(rather than 3-dimensional) that include
the state $\ket{r,r,n-1}$ (for $n \ge 1$) in addition to the 3 states 
in \eqref{states3}. See Section \ref{sec:conclusions}, \ref{sec:appTC}, and 
the text in the second paragraph below equation (1) in \cite{Tessier2003}.}.
In the basis 
$\{ \ket{\psi_1^{(n)}}, \ket{\psi_2^{(n)}}, \ket{\psi_3^{(n)}}\}$
the Hamiltonian ${\hat H}^{(n)}$ is represented by a $3 \times 3$ matrix 
\begin{equation} \label{h3matrix}
    H^{(n)} = \hbar \begin{pmatrix}
    \omega_f n & 0 & \lambda \sqrt {n+1} \\[4pt]
    0 & \omega_f n & \lambda \sqrt {n+1} \\[4pt]
    \lambda \sqrt {n+1} & \lambda \sqrt {n+1} & \omega_f n + \Delta
	\end{pmatrix} \, \, .
\end{equation}
For given $n=0,1,2, \ldots$, the energy eigenvalues and associated eigenstates of $H^{(n)}$ are
\begin{align} 
    & E_{\text{asym}\,}^{(n)} \big/ \hbar = \omega_f n \, , \label{energyasym} \\[4pt] 
    & \left\langle \vec{\psi}^{\,(n)} \Big| \text{asym}, n \right\rangle 
    = \frac{1}{\sqrt{2}} \begin{pmatrix}
    1 \\ -1 \\ 0
    \end{pmatrix} \, , \label{evasym}
\end{align}
where $\vec{\psi}^{\,(n)}$
is a vector with components $\psi_i^{(n)}$, $i=1,2,3$;
\begin{align}
    & E_{\pm}^{(n)} \big/ \hbar = \omega_f n
    + \frac{\Delta}{2} \pm \Omega_n (\Delta) \, , \label{energypm} \\[4pt]
    & \left\langle \vec{\psi}^{\,(n)} \Big| \pm, n \right\rangle 
    = \frac{1}{N_{\pm}} \begin{pmatrix}
    \pm \lambda \sqrt {n+1} \\[3pt]
    \pm \lambda \sqrt {n+1} \\[2pt]
    \Omega_n (\Delta) \pm \frac{\Delta}{2}
    \end{pmatrix} \, \, . \label{evpm}
\end{align}
The index ``asym'' in \eqref{evasym} indicates that this eigenstate 
is given by the antisymmetric superposition
\begin{equation} \label{asym}
    \ket{\text{asym}, n}
    := \frac{1}{\sqrt{2}} \left( 
    \ket{\psi_1^{(n)}} - \ket{\psi_2^{(n)}} \right)
    = \frac{1}{\sqrt{2}} \left( 
    \ket{r,g,n} - \ket{g,r,n} \right) \, .
 \end{equation}
This eigenstate does not involve the basis state $\ket{\psi_3^{(n)}} = \ket{g,g,n+1}$; 
as a result, the eigenstate \eqref{evasym} and its eigenvalue \eqref{energyasym} are 
independent of the detuning $\Delta$. In \eqref{energypm} and \eqref{evpm},
\begin{equation} \label{rabi3}
    \Omega_n(\Delta) = \sqrt{ \left( \Delta/2 \right)^2 
    + 2 \lambda^2 (n+1)}
\end{equation}
is the $\Delta$-dependent Rabi frequency for the two-atom system.
The normalization constants in \eqref{evpm} are\\ 
$N_{\pm} = \sqrt{ \left( \Omega_n \pm \Delta/2  \right)^2 + 2 \lambda^2 (n+1)}$. 

The eigenvectors in \eqref{evasym}, \eqref{evpm}
form a right-handed, orthonormal frame. 
This can be made explicit be expressing them in the form 
\begin{align} 
    & \left\langle \vec{\psi}^{\,(n)} \Big| \text{asym}, n \right\rangle     
    = \frac{1}{\sqrt{2}} \begin{pmatrix}
    1 \\ -1 \\ 0
    \end{pmatrix} \, , \, \, \label{psirot3a} \\[5pt]
    & \left\langle \vec{\psi}^{\,(n)} \Big| +, n \right\rangle 
    = \begin{pmatrix}
      \cos(\varphi_n) / \sqrt{2} \\[2pt]
      \cos(\varphi_n) / \sqrt{2} \\[2pt]    
      \sin(\varphi_n)
    \end{pmatrix} \, , \, \, 
    \label{psirot3b} \\[5pt]
    & \left\langle \vec{\psi}^{\,(n)} \Big| -, n \right\rangle 
    = \begin{pmatrix}
    -  \sin(\varphi_n) / \sqrt{2} \\[2pt]
    -  \sin(\varphi_n) / \sqrt{2} \\[2pt]    
      \cos(\varphi_n)
    \end{pmatrix} \, , \, \label{psirot3c}
\end{align}
corresponding to the column vectors of a $3 \times 3$ rotation matrix
\begin{equation} \label{R3}
    R = \begin{pmatrix}
     1/\sqrt{2} &   \cos(\varphi_n) / \sqrt{2} & -  \sin(\varphi_n) / \sqrt{2} \\[4pt]
    -1/\sqrt{2} &   \cos(\varphi_n) / \sqrt{2} & -  \sin(\varphi_n) / \sqrt{2} \\[4pt]
    0 &   \sin(\varphi_n) &   \cos(\varphi_n)
	\end{pmatrix}
\end{equation}
with an angle
\begin{equation} \label{varphi}
    \varphi_n(\Delta) = \tan^{-1} \left( 
    \frac{\Omega_n (\Delta) + \Delta/2}{\lambda \sqrt{2(n+1)}}
    \right) \, \, .
\end{equation}

\begin{figure}[t]
  \centering
  \includegraphics[width=0.68\textwidth]{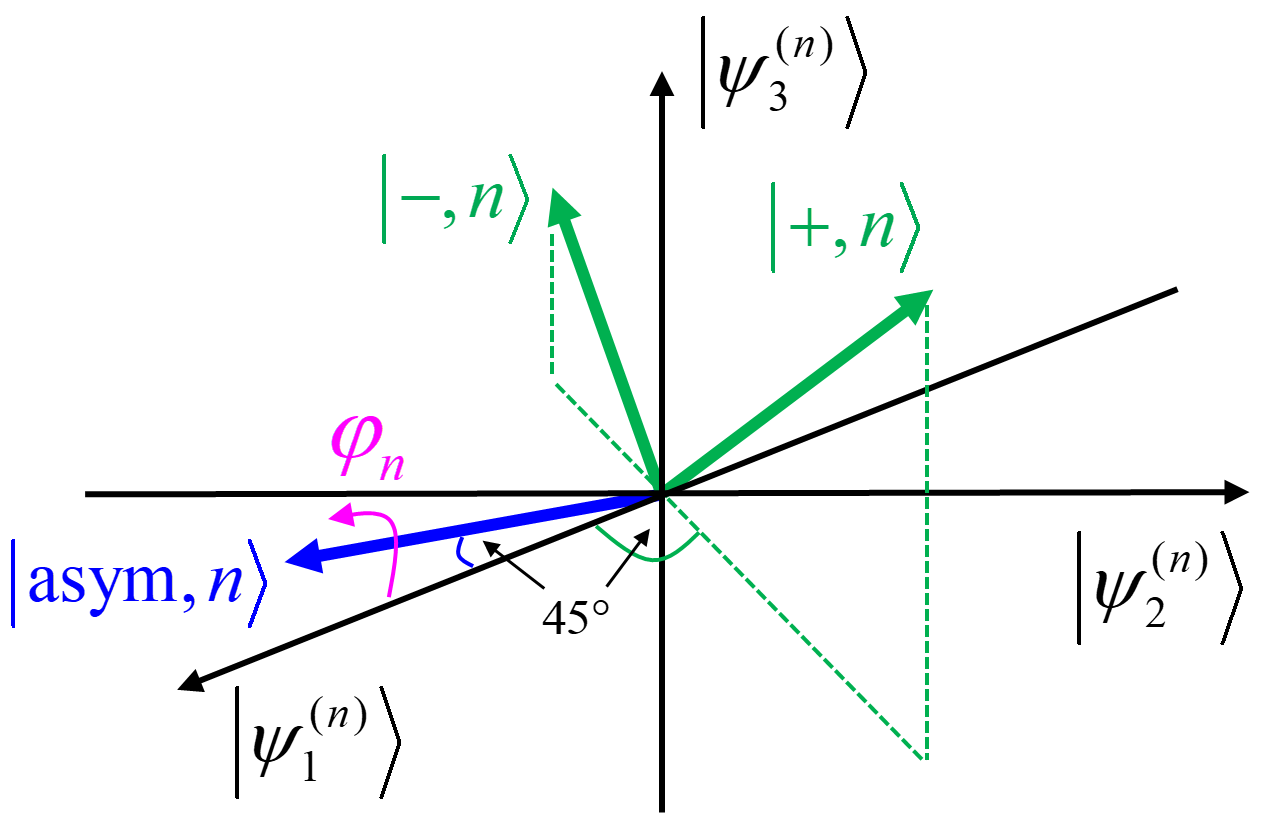}
  \caption{Rotation of the eigenvectors 
  $\ket{\pm, n}$ (bold green arrows) by the angle 
  $\varphi_n(\Delta)$ (magenta) about the fixed axis given by the eigenvector
  $\ket{\text{asym}, n}$ (bold blue arrow). 
  The rotation is described by the rotation matrix $R$ in \eqref{R3}.
  The frame is spanned by the states 
  $\ket{\psi_i^{(n)}}$, $i=1,2,3$,
  in \eqref{states3}. 
  The eigenvectors $\ket{\pm, n}$ rotate
  within the plane perpendicular to $\ket{\text{asym}, n}$ spanned by 
  $\ket{\text{sym}, n} = \tfrac{1}{\sqrt{2}} 
  ( \ket{\psi_1^{(n)}} + \ket{\psi_2^{(n)}} )$
  and $\ket{\psi_3^{(n)}}$ as indicated by the dashed lines. 
  Shown is the configuration of 
  $\ket{\pm, n}$ for zero detuning 
  ($\Delta=0$) for which $\varphi_n(0) = \pi/4 = 45^{\circ}$ and $R$ 
  is given by \eqref{R3zero}.} 
  \label{fig:rotation}
\end{figure}

\vspace*{2mm}
\noindent
Geometrically, $R$ in \eqref{R3} describes a rotation of the eigenvectors 
$\ket{\pm, n}$ by the angle $\varphi_n(\Delta)$
about the fixed axis given by the $\Delta$-independent eigenvector
$\ket{\text{asym}, n}$ (see Figure \ref{fig:rotation}).
%

The rotation matrix $R$ in \eqref{R3} generates the basis transformation 
$\{ \ket{\psi_1^{(n)}}, \ket{\psi_2^{(n)}}, \ket{\psi_3^{(n)}} \}
\to \left\{ \ket{ \text{asym}, n }, \ket{+,n}, \ket{-,n} \right\}$ 
by the relations
\begin{align} \label{psiton3}
& \ket{\text{asym},n} = \sum\limits_{i=1,2,3} R_{i,\text{asym}} \ket{\psi_i^{(n)}} \, , \\[2pt]
& \ket{\pm,n} = \sum\limits_{i=1,2,3} R_{i,\pm} \ket{\psi_i^{(n)}} \,  .
\end{align}
The inverse transformation is given by
\begin{equation} \label{ntopsi3} 
    \ket{\psi_i^{(n)}} = R_{\text{asym},\,i}^{T} \ket{\text{asym},n}
    + R_{+,\,i}^{T} \ket{+,n} + R_{-,\,i}^{T} \ket{-,n} \, , \, \, i=1,2,3 \, ,
\end{equation}
where $R^T$ is the transpose of $R$ in \eqref{R3}.
In view of the discussion in Section \ref{sec:number}, it is useful to display 
\eqref{ntopsi3} in terms of components: 
\begin{align} 
\ket{\psi_1^{(n)}}  &= \frac{1}{\sqrt{2}} \ket{\text{asym},n}
+ \frac{1}{\sqrt{2}} \cos(\varphi_n) \ket{+,n} 
- \frac{1}{\sqrt{2}} \sin(\varphi_n) \ket{-,n} \label{comp1} \\[3pt]
\ket{\psi_2^{(n)}}  &= - \frac{1}{\sqrt{2}} \ket{\text{asym},n}
+ \frac{1}{\sqrt{2}} \cos(\varphi_n) \ket{+,n}
- \frac{1}{\sqrt{2}} \sin(\varphi_n) \ket{-,n} \label{comp2} \\[2pt]
\ket{\psi_3^{(n)}}  &= \sin(\varphi_n) \ket{+,n} 
+ \cos(\varphi_n) \ket{-,n} \, \, , \label{comp3}
\end{align}
with $\varphi_n(\Delta)$ in \eqref{varphi}.

For zero detuning ($\Delta  = 0$) the eigenvector in \eqref{evasym}
and its eigenvalue in \eqref{energyasym} remain unchanged while the 
remaining expressions simplify to 
\begin{equation} \label{zero31}
    \Omega_n(0) = \lambda \sqrt{2(n+1)} \, , \, \,  \varphi_n(0) 
    = \frac{\pi}{4} \, \, , 
\end{equation}
\begin{equation} \label{zero32}
    E_\pm^{(n)} \big/ \hbar = \omega_f n \pm \lambda \sqrt{2(n+1)} \, \, ,
\end{equation}
\begin{equation} \label{zero33}
    \left\langle \vec{\psi}^{\,(n)} \Big| \pm, n \right\rangle 
    = \begin{pmatrix}
    \pm 1/2 \\[2pt] \pm 1/2 \\[2pt] 1/\sqrt{2}
    \end{pmatrix} \, \, ,
\end{equation}
\begin{equation} \label{R3zero}
    R = \begin{pmatrix}
     1/\sqrt{2} & 1/2 & -1/2 \\[3pt]
    -1/\sqrt{2} & 1/2 & -1/2 \\[3pt]
    0 & 1/\sqrt{2} & 1/\sqrt{2} 
	\end{pmatrix} \, \, .
\end{equation}
Thus, for $\Delta  = 0$ the states $\ket{\pm,n}$ are given by 
(up to a global phase factor of $- 1$ for $\ket{-,n}$)
\begin{equation} \label{pm30}
    \ket{\pm,n} 
    = \frac{1}{\sqrt{2}} \left( \ket{\text{sym}, n}
    \pm \ket{\psi_3^{(n)}} \right)
    = \frac{1}{\sqrt{2}} \left( \ket{\text{sym}, n}
    \pm \ket{g,g,n+1} \right) \, ,
\end{equation}
where (cf.~\eqref{asym}, and see Figure \ref{fig:rotation})
\begin{equation} \label{sym}
    \ket{\text{sym}, n}
    := \frac{1}{\sqrt{2}} \left( 
    \ket{\psi_1^{(n)}} 
    + \ket{\psi_2^{(n)}} \right)
    = \frac{1}{\sqrt{2}} \left( 
    \ket{r,g,n} + \ket{g,r,n} \right) \, .
 \end{equation}
The state in \eqref{sym} corresponds to the state $\ket{R}$ in \eqref{RN} for $N=2$
in the absence of the exponential prefactors $e^{i {\bf k} \cdot {\bf x}_j}$
(cp.~\eqref{lambda}) and for fixed photon number $n$.


\section{Number states of the field} 
\label{sec:number}

In this section, we discuss the dynamics of the system in the subspace 
$\mathbb{H}_n = \text{span} \{ \ket{\psi_1^{(n)}}, 
\ket{\psi_2^{(n)}}, \ket{\psi_3^{(n)}} \}$ (see \eqref{states3})
for given $n=0,1,2,\ldots$, which is described by the $3 \times 3$ 
matrix $H^{(n)}$ in \eqref{h3matrix}.  
For a given initial state at $t=0$,
\begin{equation} \label{init3}
\ket{\psi^{(n)}(0)} = \mu_0^{(n)} \ket{\psi_1^{(n)}}
+ \nu_0^{(n)} \ket{\psi_2^{(n)}} + \xi_0^{(n)} \ket{\psi_3^{(n)}} \, ,
\end{equation}
the dynamics of the system is determined by the time evolution
\begin{equation} \label{evol3}
    \ket{ \psi^{(n)}(t) } = \exp\left(-\frac{i}{\hbar} \hat{H} t \right)
    \ket{ \psi^{(n)}(0) }
    =  \mu^{(n)}(t) \ket{ \psi_1^{(n)} } + \nu^{(n)}(t) \ket{ \psi_2^{(n)} }	
    + \xi^{(n)}(t) \ket{ \psi_3^{(n)} }
\end{equation}
with $\hat{H}$ in \eqref{ham2} and
$\mu^{(n)}(t) = \langle \psi_1^{(n)} | \psi^{(n)}(t) \rangle$,
$\nu^{(n)}(t) = \langle \psi_2^{(n)} | \psi^{(n)}(t) \rangle$,
$\xi^{(n)}(t) = \langle \psi_3^{(n)} | \psi^{(n)}(t) \rangle$.
The latter relations can be combined in vector form
\begin{equation} \label{coeffvec}
    \left\langle \vec{\psi}^{\,(n)} \Big| \psi^{(n)}(t) \right\rangle
    = \begin{pmatrix}
    \mu^{(n)}(t) \\[2pt] \nu^{(n)}(t) \\[2pt] \xi^{(n)}(t)
    \end{pmatrix} \, \, .
\end{equation}
Depending on the initial state $\ket{\psi^{(n)}(0)}$ in \eqref{init3} 
the time evolution $\ket{\psi^{(n)}(t)}$ in \eqref{evol3} can be grouped in three 
general cases A – C discussed in the following subsections.  


\subsection{Case A: 
\texorpdfstring{$\ket{\psi_A^{(n)}(0)} = \ket{\text{asym}, n}$}{ }}
\label{sec:caseA}

with $\ket{ \text{asym},n }
= ( \ket{\psi_1^{(n)}} - \ket{\psi_2^{(n)}} ) / \sqrt{2}$ defined in \eqref{asym}.
This initial state corresponds to a maximally entangled Bell state of the two atoms and is an
eigenstate of $\hat H$ in \eqref{ham2} with eigenvalue $E_{\text{asym}}^{(n)} = \hbar \omega_f n$
in \eqref{energyasym}. This atomic state is a dark state that does not couple to the field. Thus, 
\begin{equation} \label{psiA}
    \ket{\psi_A^{(n)} (t)} = 
    \exp \left( - i \omega_f n t \right) \ket{ \text{asym},n }
\end{equation}
which implies for the coefficients in \eqref{evol3}
\begin{align}
    \mu_A^{(n)}(t) &= - \nu_A^{(n)}(t) 
    = \frac{1}{\sqrt{2}} \exp \left( - i \omega_f n t \right) \label{munuA} \\[2pt]
    \xi_A^{(n)}(t) & = 0 \, . \label{xiA}
\end{align}
The probability 
$P_A^{(n)}(\text{asym},t)$ to find the system in the state 
$\ket{\text{asym},n}$ is equal to 1 for all times $t>0$. 

For the cases B and C discussed below, we quote the results for the 
coefficients $\mu^{(n)}(t)$, $\nu^{(n)}(t)$, $\xi^{(n)}(t)$ in \eqref{evol3}
and the corresponding probabilities, leaving details to
\ref{sec:appA} and \ref{sec:appB}, respectively. 
If not stated otherwise, first the result for general 
detuning $\Delta$ is given, followed by the result for $\Delta=0$. 

\begin{figure}[t]
  \centering
  \includegraphics[width=0.68\textwidth]{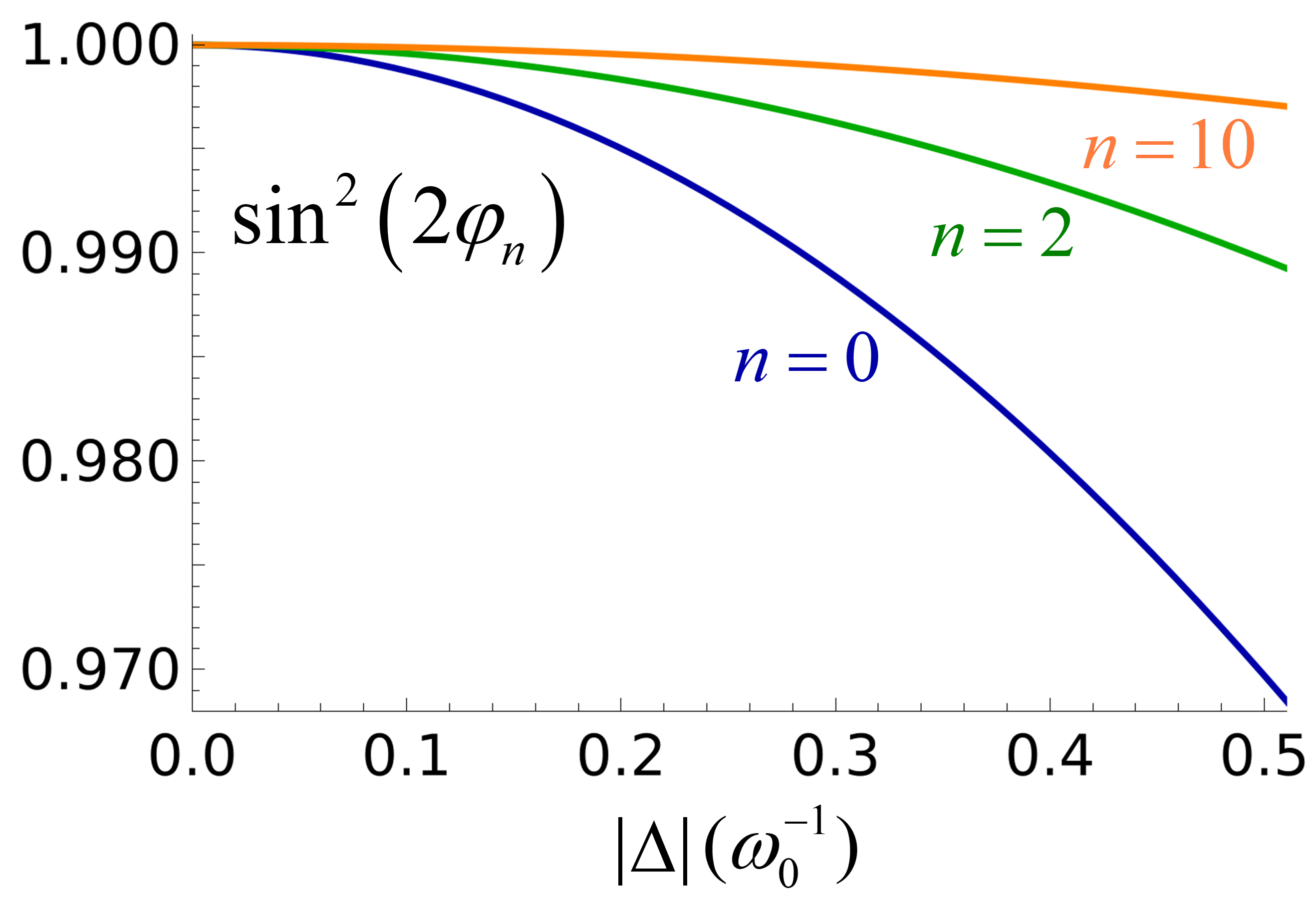}
  \caption{Amplitude $\sin^2(2 \varphi_n)$ of $P_B^{(n)}(\text{sym},t)$ in \eqref{PBsym},
  with $\varphi_n(\Delta)$ in \eqref{varphi}, as a function of the detuning $\Delta$ in units of 
  $\omega_0^{-1}$ for $\lambda  = \omega_0$ and $n=0$ (blue line), 
  2 (green), and 10 (orange).} 
  \label{fig:amplitude}
\end{figure}


\vspace{0.5cm}

\subsection{Case B: 
\texorpdfstring{$\ket{\psi_B^{(n)}(0)}
= \ket{\psi_3^{(n)}}  = \ket{g,g,n+1}$}{ }} 
\label{sec:caseB}

In this case, the two atoms are initially in their ground states 
and are exposed to the field in the photon number state $n+1$ at $t=0$. 
The coefficients in \eqref{evol3} are given by (see \ref{sec:appsubB})
    \begin{align}
    \mu_B^{(n)}(t) &= \frac{1}{\sqrt{2}} (-i) \sin(2 \varphi_n)
    \exp \Big[ -i \hspace{1pt} \Big( \omega_f n + \tfrac{\Delta}{2} \Big) \hspace{1pt} t \Big] 
    \sin(\Omega_n t) \label{B1} \\[2pt]
    &= \frac{1}{\sqrt{2}} (-i) \exp (-i \omega_f n t) 
    \sin \Big( \lambda \sqrt{2 (n+1)} \, t \Big) \, , \, \, \Delta = 0 \label{B2} \\[4pt]
    \nu_B^{(n)}(t) &= \mu_B^{(n)}(t) \label{B3} \\[4pt]
    \xi_B^{(n)}(t) &= \exp \Big[ -i \hspace{1pt} \Big( \omega_f n + \tfrac{\Delta}{2} \Big) \hspace{1pt} t \Big] 
    \left[ \cos^2(\varphi_n) \exp(i \Omega_n t)
         + \sin^2(\varphi_n) \exp(-i \Omega_n t) \right] \label{B4} \\[2pt]           
    &= \exp \Big( -i \omega_f n t \Big) 
    \cos \Big( \lambda \sqrt{2 (n+1)} \, t \Big) \, , \, \, \Delta = 0 \, \, , \label{B5}
    \end{align} 
with $\varphi_n(\Delta)$ in \eqref{varphi}. 
Since $\nu_B^{(n)}(t) = \mu_B^{(n)}(t)$ the interaction with the 
field generates the maximally entangled Bell state
$\ket{\text{sym},n}$ for the two atoms in \eqref{sym}. The probability 
$P_B^{(n)}(\text{sym},t)$ to find the system in the state 
$\ket{\text{sym},n}$ results as, using
\begin{equation} \label{sigma}
    \ket{ \psi_B^{(n)}(t) } = \sigma_B^{(n)}(t) \ket{ \text{sym},n }
    + \xi_B^{(n)}(t) \ket{ \psi_3^{(n)} } 	
\end{equation}
with $\sigma_B^{(n)}(t) := \sqrt{2} \, \mu_B^{(n)}(t)$,
\begin{align}
    P_B^{(n)}(\text{sym},t) &= \left| \sigma_B^{(n)}(t) \right|^2
    = \sin^2(2 \varphi_n) \sin^2(\Omega_n t) \label{PBsym} \\[2pt]
    &= \sin^2 \Big( \lambda \sqrt{2 (n+1)} \, t \Big) \, , \, \, \Delta = 0 \label{PBsym0} \, \, .
\end{align}
The probability to find the system in the complementary 
state $\ket{ \psi_3^{(n)} }$ is given by (see \ref{sec:appsubB})
$P_B^{(n)} \big( \ket{ \psi_3^{(n)} } , t \big) = 1 - P_B^{(n)}(\text{sym},t)$.

Thus, for the initial state $\ket{ \psi_B^{(n)} (0) } = \ket{ \psi_3^{(n)} } 
= \ket{g,g,n+1}$ the system remains in the subspace
$\text{span} \{ \ket{\text{sym},n}, \ket{ \psi_3^{(n)} } \}$
for all times $t > 0$ and oscillates with angular frequency $2 \Omega_n(\Delta)$ 
($= 2 \lambda \sqrt{2(n+1)}$ for $\Delta = 0$) between $\ket{ \psi_3^{(n)} }$
and $\ket{\text{sym},n}$. 
Since $\ket{\text{sym},n}$ in \eqref{sym}
corresponds to a maximally entangled Bell state for the two atoms, the 
interaction of the quantized field with two atoms initially in the ground state generates an 
entangled state of the atoms that is orthogonal to the entangled state $\ket{\text{asym},n}$ of case A
(Section \ref{sec:caseA}).

Due to the amplitude $\sin^2(2 \varphi_n)$ in \eqref{PBsym}, for $\Delta \ne 0$ the value of
$P_B^{(n)}(\text{sym},t)$ at the maxima is slightly less than 1. For example, for $\lambda  = \omega_0$,
$n = 0$, and $\Delta = 0.5 \, \omega_0$, one has $\sin^2(2 \varphi_n) \simeq 0.97$. 
This effect becomes even smaller for increasing $n$ (Figure \ref{fig:amplitude}).

In view of Rydberg systems as potential building blocks of atom-light quantum interfaces, an important quantity is 
the entanglement between the two-atom system and the quantized field
\cite{Kuzmich2003,Reiserer2015,Tiarks2019}.
Assuming zero detuning for simplicity ($\Delta=0)$, \eqref{sigma} and 
\eqref{B2}, \eqref{B5} imply that at times $t$ for which  
$\lambda \sqrt{2 (n+1)} \, t = \pi/4$ the state of the combined system 
(atoms + field) is given by
\begin{equation} \label{Cent}
    \ket{ \psi_B^{(n)}(t) } = \frac{\exp (-i \omega_f n t)}{\sqrt{2}} 
    \left( -i \ket{ \text{sym},n } + \ket{ \psi_3^{(n)}} \right)
\end{equation}    
with $\ket{ \text{sym},n }$ in \eqref{sym} and 
$\psi_3^{(n)} = \ket{g,g,n+1}$ in \eqref{states3}.
The state \eqref{Cent} corresponds to a maximally entangled state
between the qubit encoded by the two-atom system ($\alpha$) as
$\left\{\ket{g,g}_{\alpha}, \ket{\text{sym}}_{\alpha} \right\}$ 
and the qubit encoded by the number states of the field ($f$) as  
$\left\{\ket{n}_f, \ket{n+1)}_f \right\}$.
If the photons of the field, after interacting with the two-atom system ($\alpha$)
resulting in \eqref{Cent}, are brought in contact with a second, 
separate two-atom system ($\beta$) 
initially prepared in the state $\ket{g,g}_{\beta}$, then the second two-atom system 
will evolve to a superposition of $\ket{g,g}_{\beta}$ and $\ket{\text{sym}}_{\beta}$
which depends on the photon qubit $\left\{\ket{n}_f, \ket{n+1)}_f \right\}$. 
That is, the second two-atom system ($\beta$) will be entangled with the first ($\alpha$)
via the photons of the field. Although this scheme may be challenging to implement 
experimentally since it requires precise control of the photon number $n$, it illustrates 
how systems described by the extended Jaynes-Cummings model \eqref{ham} can in principle be used
to entangle stationary atomic systems via photons. 
In Section \ref{sec:entanglement} we discuss different types of entanglements 
between the atoms and a quantum single-mode coherent field, which does not require 
precise control of individual photon number states $\ket{n}$.


\subsection{Case C: 
\texorpdfstring{$\ket{\psi_C^{(n)}(0)}
= \ket{\psi_1^{(n)}}  = \ket{r,g,n}$}{ }} 
\label{sec:caseC}

In this case, one of the atoms is initially excited while the other
is in the ground state, and the two atoms are exposed to the field
in the photon number state $\ket{n}$ at $t=0$.
Because the Hamiltonian \eqref{ham} is symmetric under atom exchange,
without restriction we assume that atom $a$ is 
excited and atom $b$ is in the ground state.
The coefficients in \eqref{evol3} are given by (see \ref{sec:appB})
    \begin{align}
    \begin{matrix}
    \mu_C^{(n)}(t) \\[3pt] \nu_C^{(n)}(t) 
    \end{matrix} 
    \, \Bigg\} &= \frac{1}{2} \exp (-i \omega_f n t) 
    \Big[\pm 1 + \exp(-i \tfrac{\Delta}{2} t ) \Big]
    \Big[\sin^2(\varphi_n) \exp(i \Omega_n t)
    + \cos^2(\varphi_n) \exp(-i \Omega_n t) \Big] \label{C1} \\
    &= \frac{1}{2} \exp (-i \omega_f n t) 
    \Big[\pm 1 + \cos \Big( \lambda \sqrt{2 (n+1)} \, t \Big) \Big]
    \, , \, \, \Delta = 0 \label{C2} \\[5pt]
    \xi_C^{(n)}(t) &= \mu_B^{(n)}(t) \text{ in \eqref{B1}} \, , \label{C3}
    \end{align}
where the $(+)$ sign in \eqref{C1}, \eqref{C2} belongs to $\mu_C^{(n)}(t)$
and the $(-)$ sign belongs to $\nu_C^{(n)}(t)$. The angle
$\varphi_n(\Delta)$ is given in \eqref{varphi}. 
For the probabilities of $\ket{\psi_1^{(n)}}$ and
$\ket{\psi_2^{(n)}}$ (given by 
$|\hspace{1pt} \mu_C^{(n)}(t) |^2$ and $| \nu_C^{(n)}(t)|^2$, respectively)
we only quote explicit formulae 
for zero detuning ($\Delta=0$) for simplicity; results 
for $\Delta \ne 0$ are shown in Figure \ref{fig:Cd2} below. 
\begin{align} 
    P_C^{(n)} \big( \ket{\psi_{1,2}^{(n)}}, t \big)
    &= \frac{1}{4} \left[ 1 \pm \cos \Big( \lambda \sqrt{2 (n+1)} \, t \Big) \right]^2 \, , 
    \, \, \Delta = 0 \label{PC1} \\[4pt]
    P_C^{(n)} \big( \ket{\psi_3^{(n)}}, t \big) 
    &= \frac{1}{2} P_B^{(n)}(\text{sym},t) \, \, , \,
    \text{see \eqref{PBsym}} \, \, . \label{PC2}
\end{align}
In \eqref{PC1} the $(+)$ sign belongs to $\ket{\psi_1^{(n)}}$
and the $(-)$ sign belongs to $\ket{\psi_2^{(n)}}$.

\begin{figure}[t]
  \centering
  \includegraphics[width=0.68\textwidth]{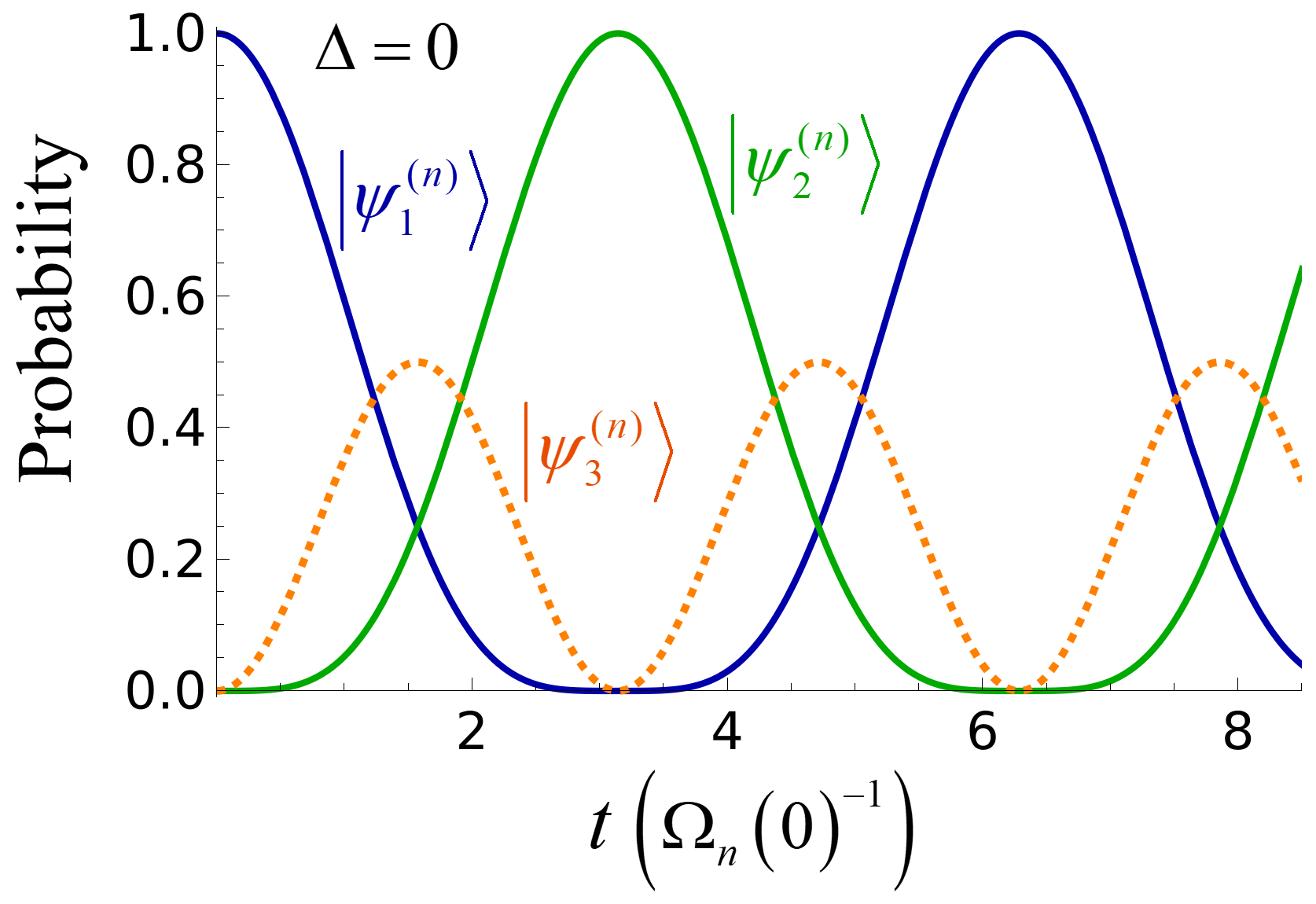}
  \caption{Time-dependence of the probabilities 
  $P_C^{(n)} \big( \ket{ \psi^{(n)}}, t \big)$ in 
  \eqref{PC1}, \eqref{PC2} for zero detuning ($\Delta  = 0$)
  in units of 
  $\Omega_n(0)^{-1} = \big( \lambda \sqrt{2(n+1)} \, \big)^{-1}$,
  for $\ket{ \psi_1^{(n)}}  = \ket{r,g,n}$ (blue solid line),
  $\ket{ \psi_2^{(n)}}  = \ket{g,r,n}$ (green solid line), 
  and $\ket{ \psi_3^{(n)}}  = \ket{g,g,n+1}$ (orange dotted line).} 
  \label{fig:Cd0}
\end{figure}

Figure \ref{fig:Cd0} shows the time-dependence of the probabilities in
\eqref{PC1}, \eqref{PC2} for $\Delta  = 0$ in units of 
$\Omega_n(0)^{-1} = \big( \lambda \sqrt{2(n+1)} \, \big)^{-1}$. 
The probabilities of the states $\ket{ \psi_1^{(n)} } = \ket{r,g,n}$ and
$\ket{ \psi_2^{(n)} } = \ket{g,r,n}$
oscillate with angular frequency $\Omega_n(0)$ between 0 and 1 and remain
completely out of phase at all times. The shape of these probabilities is not purely sinusoidal.
Conversely, the probability of $\ket{ \psi_3^{(n)} } = \ket{g,g,n+1}$
has a sinusoidal shape and oscillates with angular frequency $2 \Omega_n(0)$ between 
0 and $1/2$.    

Figure \ref{fig:Cd2} shows the probabilities corresponding to Figure \ref{fig:Cd0} 
for finite detuning $\Delta  = 0.2 \, \omega_0$. In
this case, the time-dependence of the probabilities depends explicitly on
$\lambda$ and $\omega_0$ as well as on the photon number $n$; 
in Figure \ref{fig:Cd2}, we set $\lambda = \omega_0$, express the time-dependence
in units of $\omega_0^{-1}$, and show results for $n = 0$ and $n = 10$
(parts A and B). For $\Delta \ne 0$, the relative phase between 
$\ket{\psi_1^{(n)}}$ and $\ket{\psi_2^{(n)}}$
is no longer constant but oscillates with angular frequency
$\Delta$, which for $\Delta \ll \Omega_n$ results in an envelope of the waveform
that varies slowly compared to the oscillations of the probabilities; the frequency 
of the latter is an increasing function of $n$. 
Conversely, the probability of $\ket{\psi_3^{(n)}}$
retains a sinusoidal shape also for nonzero $\Delta$ and oscillates with
angular frequency $2 \Omega_n(\Delta)$ between 0 and 
$\sin^2(2{\varphi_n}) / 2$ (see \eqref{PC2}, \eqref{PBsym}, 
and Figure \ref{fig:amplitude}).

\begin{figure}[t]
  \centering
  \includegraphics[width=0.68\textwidth]{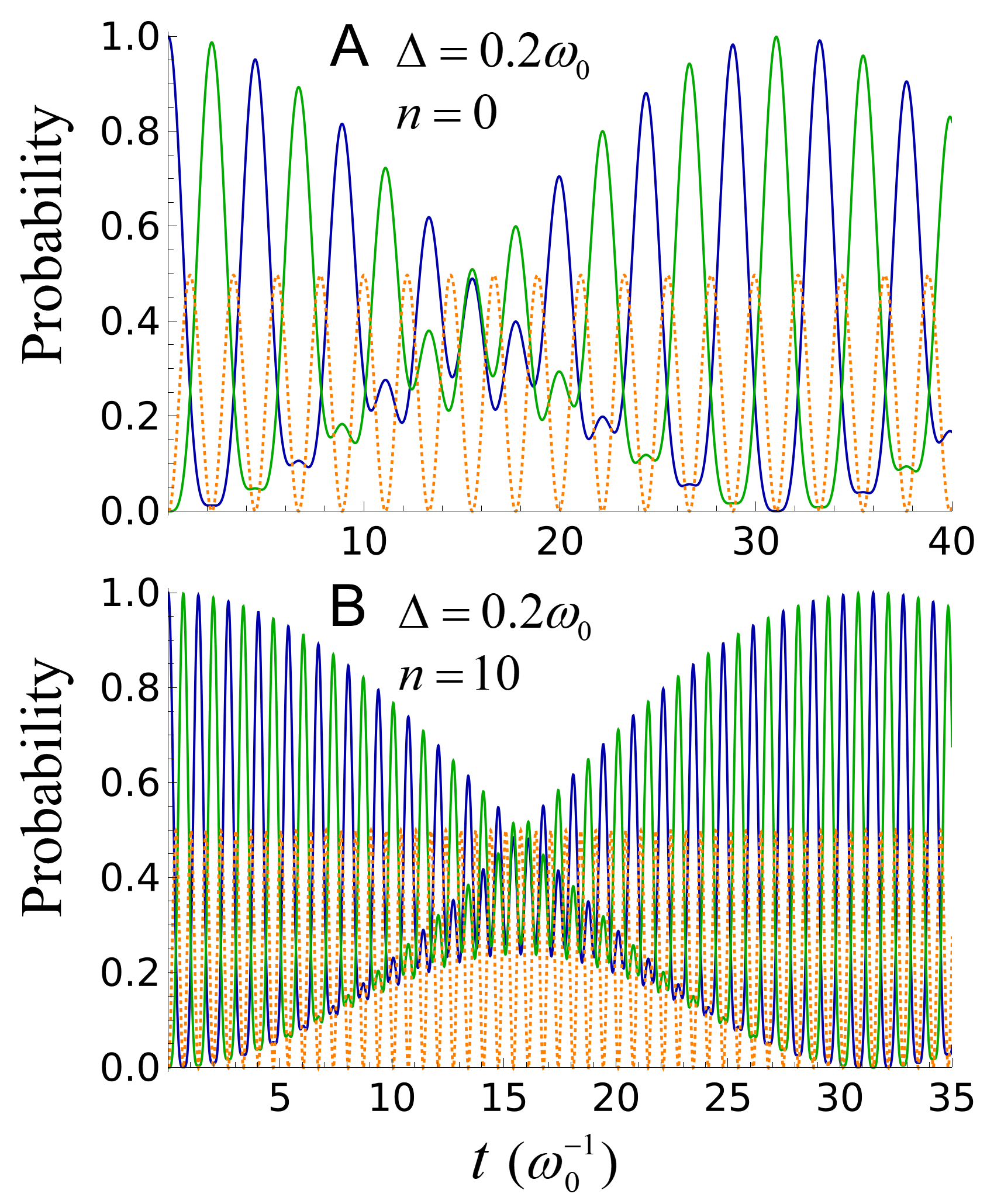}
  \caption{Time-dependence (in units of $\omega_0^{-1}$) of the probabilities 
  $P_C^{(n)}\big( \ket{\psi^{(n)}}, t \big)$ for 
  $\ket{\psi_1^{(n)}} = \ket{r,g,n}$ (blue solid line),
  $\ket{\psi_2^{(n)}} = \ket{g,r,n}$ (green solid line), 
  and $\ket{\psi_3^{(n)}} = \ket{g,g,n+1}$ (orange dotted line)
  for $\lambda = \omega_0$, detuning $\Delta  = 0.2 \, \omega_0$,
  and (A) $n=0$, (B) $n=10$.}
  \label{fig:Cd2}
\end{figure}


\section{Coherent states of the field} 
\label{sec:coherent}

In this section we consider the case that the two atoms are initially in their ground states $\ket{g}$ and the field is in a quantum single-mode coherent state
\begin{equation} \label{cs}
   \ket{\alpha}  = \exp \left( - \frac{\left| \alpha \right|^2}{2} \right)
   \sum\limits_{m=0}^{\infty} \frac{\alpha^m}{\sqrt{m!}} \ket{m} \, , \, \, t = 0 \, , 
\end{equation}
where $\ket{m}$, $m = 0,1,2,\ldots$ are photon number states of the field
discussed in Section \ref{sec:number}, and
$\alpha$ is a complex coefficient whose amplitude sets the average number of photons to
$\bar{n} = \left| \alpha \right|^2$. 
The quantum coherent state $\ket{\alpha}$ can describe the radiation field 
in an electromagnetic cavity (Figure \ref{fig:atoms}) or a propagating laser beam.
Thus, the initial state of the system is given by 
\begin{align} 
\ket{\psi^{(\alpha)}(0)} &= \ket{g,g,\alpha} 
    := \exp \left( - \frac{\left| \alpha \right|^2}{2} \right)
   \sum\limits_{m=0}^{\infty} \frac{\alpha^m}{\sqrt{m!}} \ket{g,g,m} \label{ic} \\[2pt]
   &= \exp \left( - \frac{\left| \alpha \right|^2}{2} \right) 
   \left[ \ket{g,g,0} +
   \sum\limits_{n=0}^{\infty} \frac{\alpha^{n+1}}{\sqrt{(n+1)!}} \ket{g,g,n+1} \right] 
   \, ,
\end{align}
where $\ket{g,g,n+1}$ are the basis states $\ket{\psi_3^{(n)}}$ 
of the subspaces $\mathbb{H}_n = \text{span} \{ \ket{\psi_1^{(n)}}, 
\ket{\psi_2^{(n)}}, \ket{\psi_3^{(n)}} \}$
introduced below \eqref{states3}. Thus, for the Hamiltonian
$\hat{H}$ in \eqref{ham2}, the initial 
state $\ket{\psi^{(\alpha)}(0)}$ in \eqref{ic} evolves to the state 
\begin{equation} \label{evolc} 
    \ket{ \psi^{(\alpha)}(t) } =  
    \exp \left( - \frac{\left| \alpha \right|^2}{2} \right)  
    \left[ \exp(i \omega_0 t) \ket{g,g,0} +
    \sum\limits_{n=0}^{\infty} \frac{\alpha^{n+1}}{\sqrt{(n+1)!}} 
    \ket{\psi_B^{(n)}(t)} \right] \, , \, \, t>0 \, ,
\end{equation}
with $\ket{\psi_B^{(n)}(t)}$ in \eqref{sigma}.

We now consider the probability 
$P_{\text{atoms}}^{(\alpha)} \left( \ket{\psi_{\text{atoms}}},t \right)$
to find the subsystem of the two atoms at $t > 0$ in the state 
$\ket{ \psi_{\text{atoms}} }$, 
where the initial state $\ket{\psi^{(\alpha)}(0)}$ 
of the total system (atoms + field) is given by \eqref{ic}.
Since according to \eqref{sigma} for every $n = 0,1,2,\ldots$ the state 
$\ket{\psi_B^{(n)}(t)}$ in \eqref{evolc} is a linear combination of 
$\ket{\text{sym},n}$ and $\ket{\psi_3^{(n)}}$,
the state $\ket{ \psi_{\text{atoms}} }$ 
is contained in the two-dimensional subspace
$\mathbb{H}_{\text{atoms}} := \text{span}
\{ \ket{\text{sym}}, \ket{g,g} \}$, with 
\begin{equation} \label{symatoms} 
    \ket{\text{sym}} := \frac{1}{\sqrt{2}} \left( \ket{r,g} + \ket{g,r} \right) \, .
\end{equation}
The probability 
$P_{\text{atoms}}^{(\alpha)} \left(\ket{\text{sym}},t \right)$ to find
the two atoms at $t > 0$ in the state $\ket{\text{sym}}$ in \eqref{symatoms}
is given by the matrix element
\begin{equation} \label{Psymatoms1}
    P_{\text{atoms}}^{(\alpha)} \left(\ket{\text{sym}},t \right) =
    \langle \text{sym} \, | \, \hat{\rho}_{\text{atoms}}(t) \,
    | \, \text{sym} \rangle \, ,
\end{equation}
where the reduced density operator for the atoms, $\hat{\rho}_{\text{atoms}}(t)$,
is given by the partial trace of the density operator 
\begin{equation} \label{rho}
    \hat{\rho}(t) = | \psi^{(\alpha)}(t) \rangle \langle \psi^{(\alpha)}(t) |
\end{equation}
of the total system (atoms + field) in the pure state 
$\ket{ \psi^{(\alpha)}(t) }$ 
over the field $\ket{\phi_f}$ (see \eqref{prod3}), i.e.,  
\begin{equation} \label{rhoatoms}
    \hat{\rho}_{\text{atoms}}(t) = \text{tr}_f \left\{ \hspace{1pt}\hat{\rho}(t) \right\}
    = \sum\limits_{k=0}^{\infty} \langle k \, | \, \psi^{(\alpha)}(t) \rangle
    \langle \psi^{(\alpha)}(t) \, | \, k \rangle \, ,
\end{equation}
where $\ket{k}$, $k=0,1,2,\ldots$, are the photon number states of the field. 
Combining \eqref{evolc} - \eqref{rhoatoms} yields
\begin{align} 
    & P_{\text{atoms}}^{(\alpha)} \left(\ket{\text{sym}},t \right)
    = \sum\limits_{k=0}^{\infty} \langle \text{sym}, k \,| \, \psi^{(\alpha)}(t) \rangle
    \langle \psi^{(\alpha)}(t) \, | \, \text{sym}, k \rangle \label{Psymatoms1a} \\[2pt]
    & \quad = \exp(-\bar{n}) \sum\limits_{m=1}^{\infty} \frac{\bar{n}^m}{m!}
    \left| \sigma_B^{(m-1)}(t) \right|^2 \label{Psymatoms2} \\[2pt]
    & \quad = \exp(-\bar{n}) \sum\limits_{m=1}^{\infty} \frac{\bar{n}^m}{m!}
    \sin^2 \left( \lambda \sqrt{2m} \, t \right) \, , \, \, \Delta=0 \, , \label{Psymatoms3}
\end{align}
with $\sigma_B^{(n)}(t) := \sqrt{2} \, \mu_B^{(n)}(t)$ in \eqref{sigma}.  
As in \eqref{B1}, \eqref{B2}, first the result for general detuning 
$\Delta = \omega_f - \omega_0$ is quoted in \eqref{Psymatoms2},
followed by the result for $\Delta=0$ in \eqref{Psymatoms3}.
A similar calculation yields for the probability 
to find the two atoms at $t > 0$ in the state $\ket{g,g}$
the result
\begin{align}
    & P_{\text{atoms}}^{(\alpha)} \left(\ket{g,g},t \right) =
    \langle g,g \, | \, \hat{\rho}_{\text{atoms}}(t) \, 
    | \, g,g \rangle = \exp(-\bar{n}) \left(
    1 + \sum\limits_{m=1}^{\infty} \frac{\bar{n}^m}{m!}
    \left| \xi_B^{(m-1)}(t) \right|^2 \right) \label{Pggatoms1} \\[2pt]
    & \quad = \exp(-\bar{n})
    \sum\limits_{m=0}^{\infty} \frac{\bar{n}^m}{m!}
    \cos^2 \left(\lambda \sqrt{2m} \, t \right) \, , \, \, \Delta=0 \, , \label{Pggatoms2}
\end{align}
with $\xi_B^{(n)}(t)$ in \eqref{B4}.
The normalization condition 
$P_{\text{atoms}}^{(\alpha)} \left(\ket{\text{sym}},t \right) 
+ P_{\text{atoms}}^{(\alpha)} \left(\ket{g,g},t \right) = 1$
is fulfilled due to $\left| \sigma_B^{n}(t) \right|^2 + 
\left| \xi_B^{n}(t) \right|^2 = 1$ for $n=0,1,2,\ldots$ 
(see \eqref{sigma}).

Figure \ref{fig:coherent} shows 
$P_{\text{atoms}}^{(\alpha)} \left(\ket{\text{sym}},t \right)$ in \eqref{Psymatoms1}
and $P_{\text{atoms}}^{(\alpha)} \left(\ket{g,g},t \right)$ in \eqref{Pggatoms1}
for zero detuning ($\Delta  = 0$) and 
$\bar{n} = 10$, $20$, and $50$ as functions of time $t$ 
(orange and gray lines in parts (A), (C), (E),
respectively). The oscillations of the probabilities
undergo collapses and revivals on a time scale $\sim \lambda \sqrt{\bar{n}}$, 
similarly as for a single two-level atom interacting with a quantum single-mode 
coherent field \cite{Eberly1980,Larson2021}. However, for the present two-atom system, 
the oscillations occur between $\ket{g,g}$ and the maximally entangled Bell state 
$\ket{\text{sym}}$ in \eqref{symatoms}; 
thus, the interaction of the two-atom system with a quantum single-mode 
coherent field entangles the two atoms, where the time-dependence of 
the states and the entanglement can be tuned by the average photon number
$\bar{n}$ of the coherent field. The phenomenon of entanglement 
will be discussed in more detail in the following section.


\section{Entanglement} 
\label{sec:entanglement}

\begin{figure}[h]
  \centering
  \includegraphics[width=0.68\textwidth]{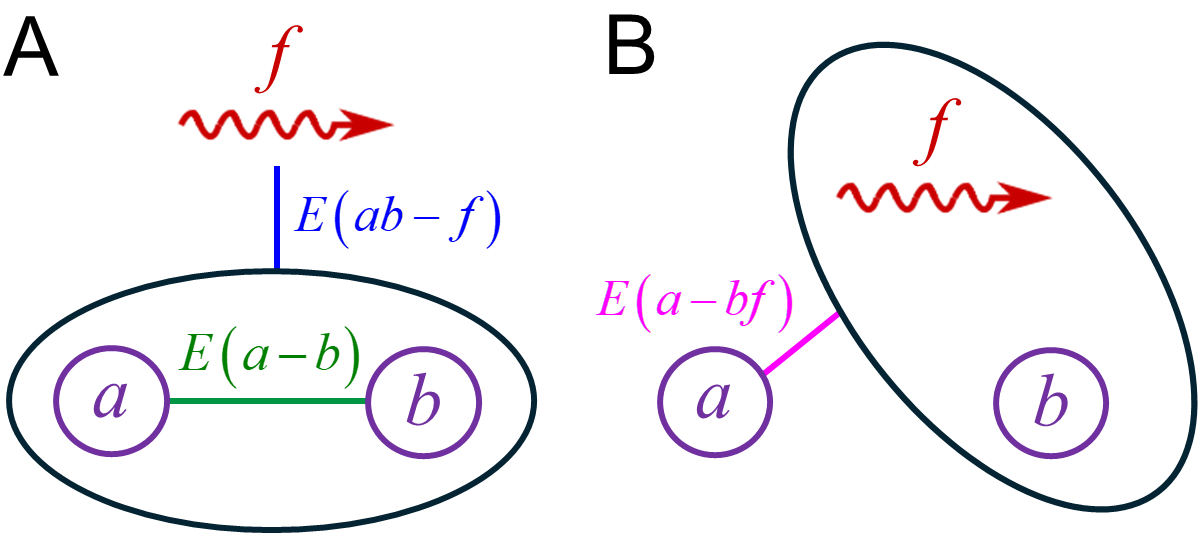}
  \caption{Entanglements between various components of the 
  total, tripartite system consisting of the two Rydberg-interacting atoms $a$, $b$ (purple) and the 
  quantized radiation field $f$ (red). The total system is assumed to be in the pure 
  state $\ket{ \psi^{(\alpha)}(t)}$ in \eqref{evolc}.
  (A) Entanglement $E(ab-f)$ (blue) between the composite two-atom system $ab$ (enclosed by the ellipse)
  and the field $f$, and entanglement of formation $E(a-b)$ (green) between the two atoms $a$, $b$.
  (B) Entanglement $E(a-bf)$ (magenta) between atom $a$ and the composite (atom $b$)-field 
  system $bf$ (enclosed by the ellipse).}
  \label{fig:entanglement}
\end{figure}

In view of the potential use of Rydberg systems for quantum state transfer 
between atoms and photons in quantum information processing
\cite{Kudryavtsev1993,Dung1994,Walther2006,Bashkirov2008,Nielsen_Chuang_2010,Chen2010,Li2013,Kumar2016,Li2019},
it is of interest 
to study the entanglement between various components of the 
total system consisting of the two Rydberg-interacting atoms $a$, $b$ 
and the quantized radiation field (see Figure \ref{fig:entanglement}).
For non-interacting atoms described by the Tavis-Cummings model, 
corresponding results were obtained in reference \cite{Tessier2003}
(see Section \ref{sec:conclusions}). 

To be specific, we assume that the field is in a quantum coherent state 
$\ket{\alpha}$ given by \eqref{cs}.  
The time evolution of the entanglement between the atoms 
and the field depends crucially on the initial condition at $t=0$. 
For example, if the atomic system is initially in the dark state 
$\ket{\text{asym}} = \frac{1}{\sqrt{2}} \left( \ket{r,g} - \ket{g,r} \right)$
(Section \ref{sec:caseA})
the total system (atoms + field) will remain a product state $\ket{\text{asym}} \otimes \ket{\alpha}$
for all times $t>0$ because the dark state does not couple to the field. 
This implies that the atoms will remain disentangled from the field. 
Conversely, if both atoms are initially in their ground state $\ket{g}$
(Section \ref{sec:caseB}) the atomic system will remain in the 
two-dimensional subspace perpendicular to $\ket{\text{asym}}$
and spanned by $\ket{g,g}$ and 
$\ket{\text{sym}} = \tfrac{1}{\sqrt{2}} \left(\ket{r,g} + \ket{g,r} \right)$
for all times $t>0$ (compare Figure \ref{fig:rotation}).
In this case, the time evolution of the total system (atoms + field) 
is given by $\ket{ \psi^{(\alpha)}(t)}$ in \eqref{evolc} and discussed in Section \ref{sec:coherent}.
In what follows, we focus on this case because 
we expect that the entanglement between the atoms and the field will be strongest 
when the atomic system is perpendicular to the dark state $\ket{\text{asym}}$. 
Results for other initial conditions, e.g., one atom excited and the other in 
the ground state, can be obtained using the results in Section \ref{sec:number}. 

\subsection{Entanglement between the composite two-atom system $ab$ and the field $f$}
\label{sec:ab-f}

Consider a composite quantum system $Q$ with Hilbert space $\mathbb{H}$
and a bipartition $A | B$ of this system into subsystems $A$ and $B$ 
with respective Hilbert spaces $\mathbb{H}_A$ and $\mathbb{H}_B$.
The Hilbert space of the composite system $Q$ is the tensor product
$\mathbb{H} = \mathbb{H}_A \otimes \mathbb{H}_B$.
A state $\ket{\psi} \in \mathbb{H}$ of the composite system is separable
across the bipartition $A | B$ if it can be represented as a product state
$\ket{\psi} = \ket{\psi}_A \otimes \ket{\psi}_B$ with $\ket{\psi}_A \in \mathbb{H}_A$
and $\ket{\psi}_B \in \mathbb{H}_B$. Otherwise, $\ket{\psi}$ is inseparable, 
or entangled, across the bipartition $A | B$ \cite{Nielsen_Chuang_2010,Horodecki2009}.

If the composite system is in a pure state $\ket{\psi}$ with density operator $\hat{\rho} = \ketbra{\psi}{\psi}$
a quantitative measure for the entanglement $E(A-B)$ between the two subsystems $A$ and $B$ is
the von Neumann entropy of either of the two subsystems $A$ and $B$
\cite{Nielsen_Chuang_2010,Horodecki2009}:
\begin{equation} \label{vne}
E(A-B) = S_A = - \text{tr}\left\{ \hspace{1pt} \hat{\rho}_A \log_2 \hat{\rho}_A \right\}
= S_B = - \text{tr} \left\{ \hspace{1pt} \hat{\rho}_B \log_2 \hat{\rho}_B \right\} \, ,
\end{equation}
where the reduced density operator $\hat{\rho}_A$ of subsystem $A$ is the partial trace of 
$\hat{\rho}$ over subsystem $B$, and similarly $\hat{\rho}_B$ is the partial trace of 
$\hat{\rho}$ over subsystem $A$.
$E(A-B)$ is zero if and only if $\hat{\rho}_A$ and $\hat{\rho}_B$ represent pure states 
$\ket{\psi}_A$ and $\ket{\psi}_B$ of subsystems $A$ and $B$, respectively, which implies that 
the composite system is in a product state 
$\ket{\psi} = \ket{\psi}_A \otimes \ket{\psi}_B$ and hence the subsystems are not entangled. 
For any entangled state, $E(A-B)$ is larger than zero. 

\begin{figure}[H]
    \centering
    \begin{minipage}{0.49\textwidth}
    \centering
    \includegraphics[width=1\textwidth]{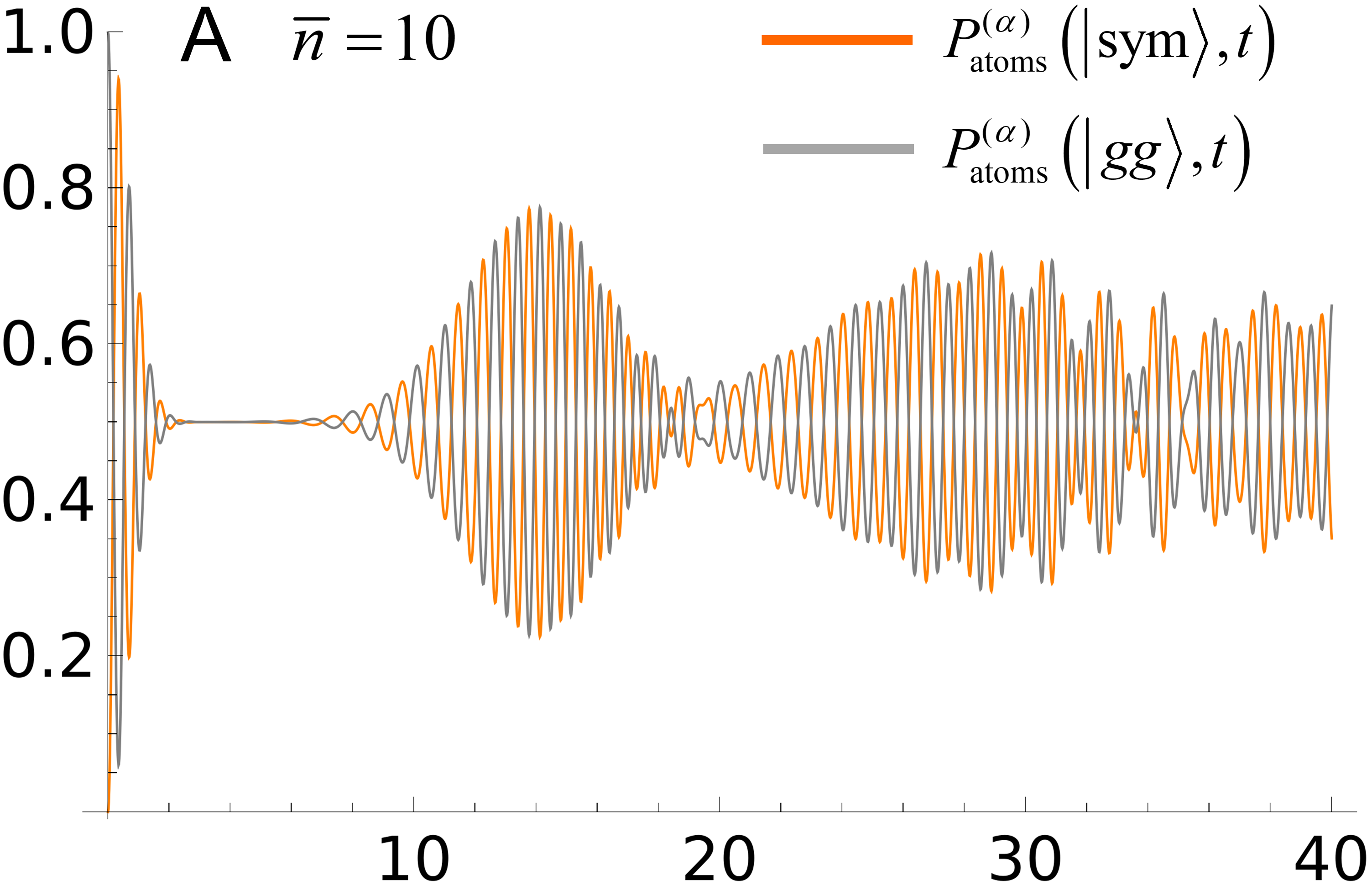}
    \end{minipage}
    \hfill
    \begin{minipage}{0.49\textwidth}
    \centering
    \includegraphics[width=1\textwidth]{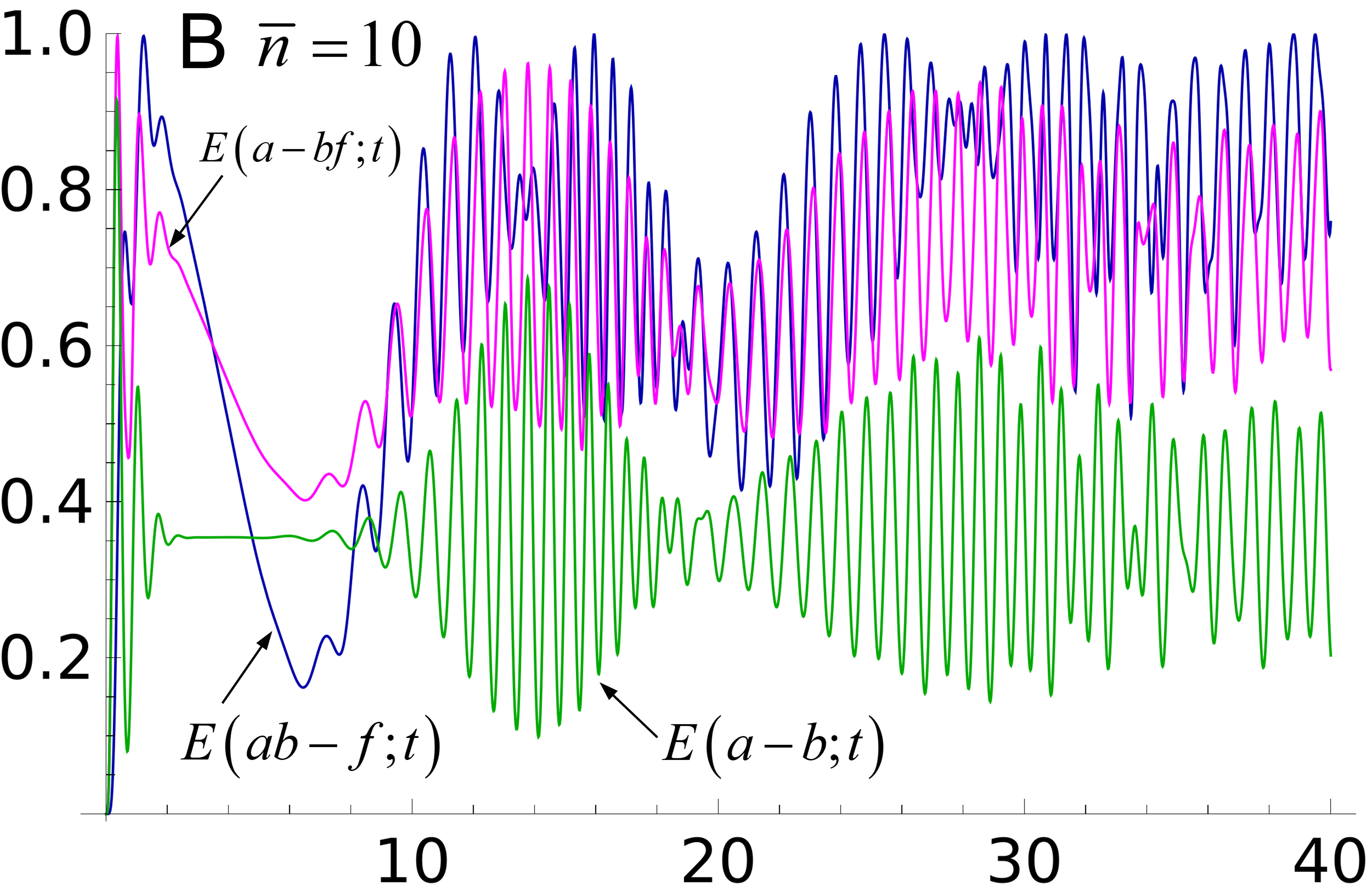}
    \end{minipage}
\par\bigskip
\begin{minipage}{0.49\textwidth}
    \centering
    \includegraphics[width=1\textwidth]{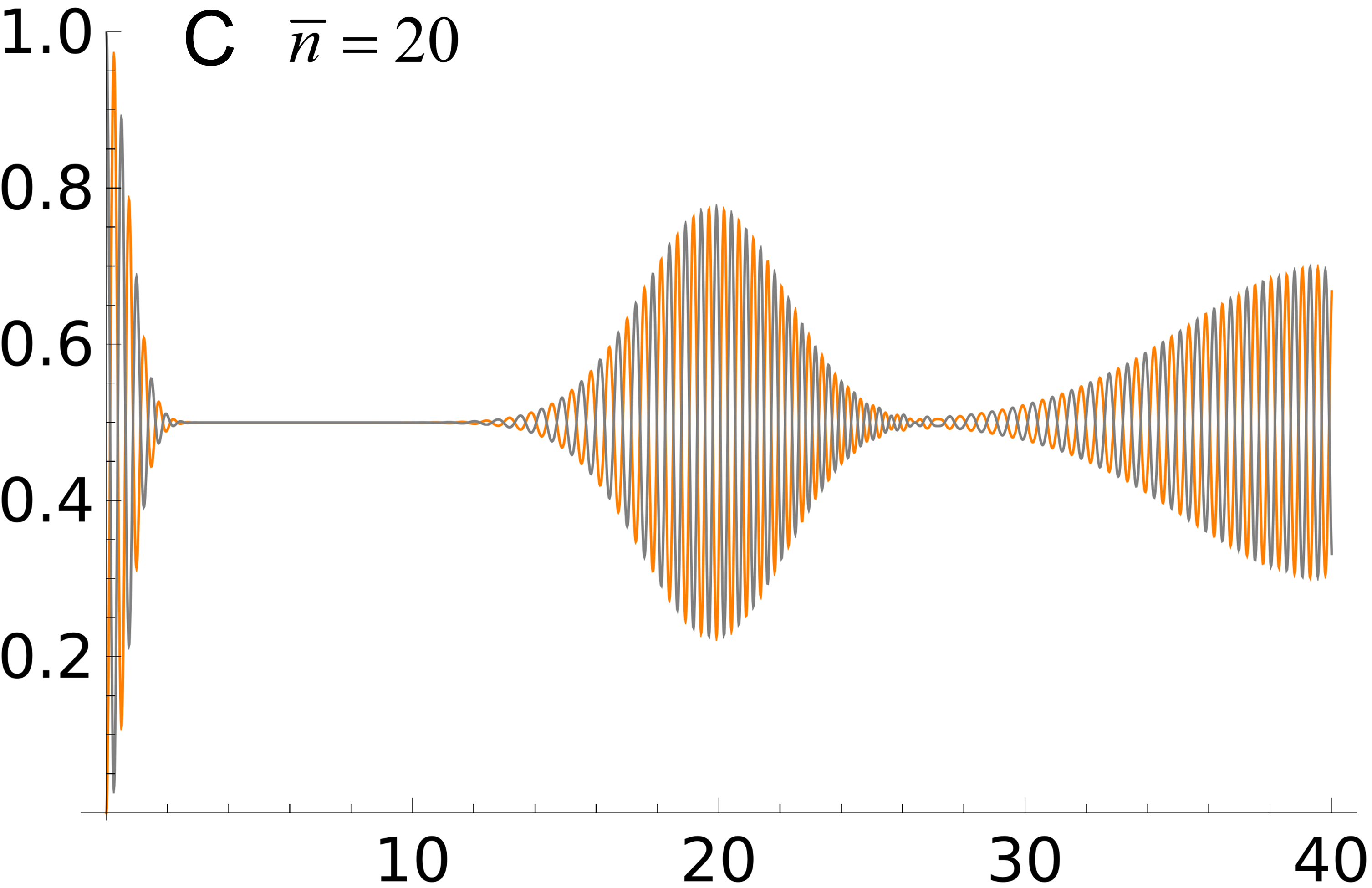}
    \end{minipage}
    \hfill
    \begin{minipage}{0.49\textwidth}
    \centering
    \includegraphics[width=1\textwidth]{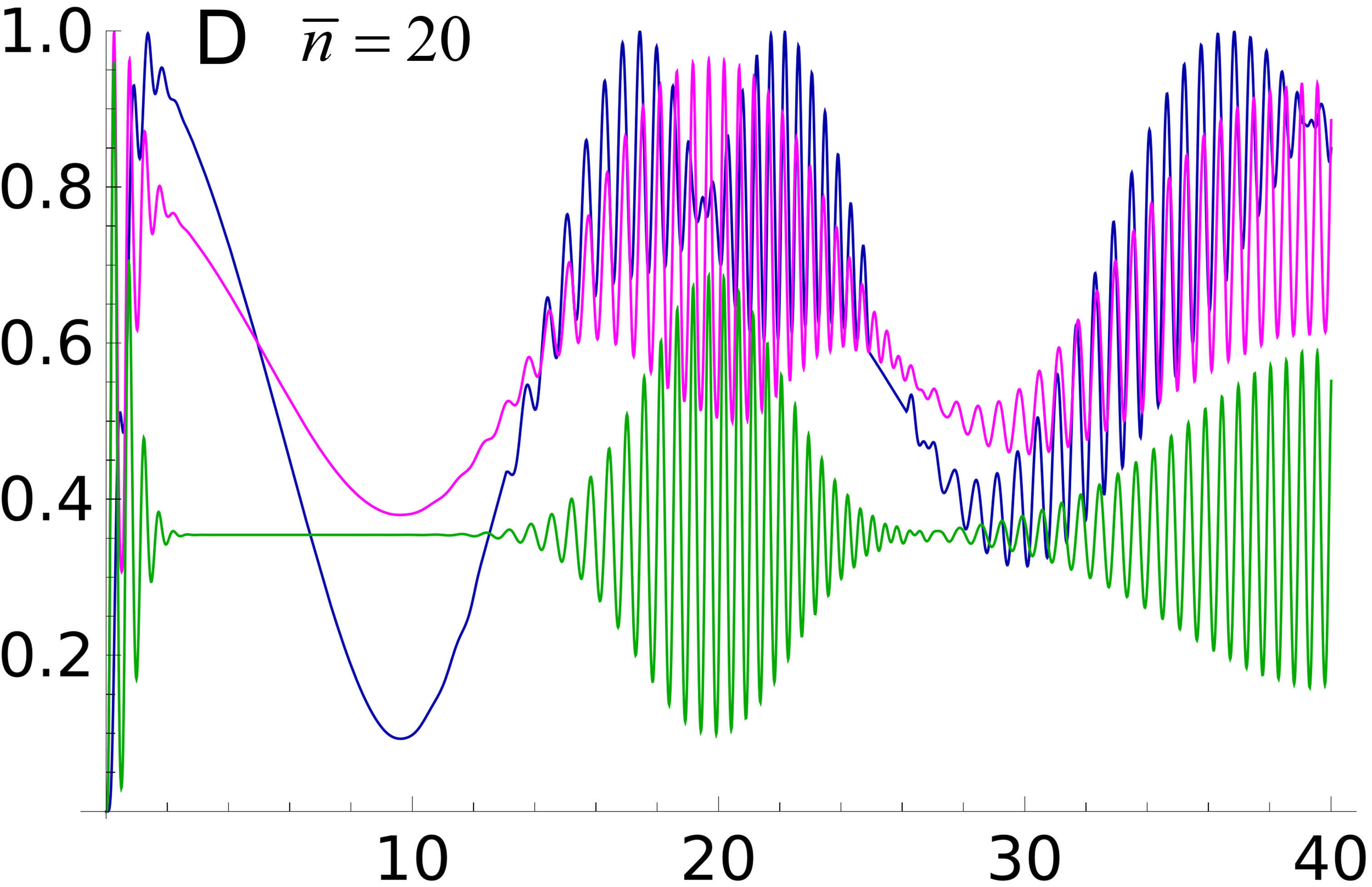}
    \end{minipage}
\par\bigskip
\begin{minipage}{0.49\textwidth}
    \centering
    \includegraphics[width=1\textwidth]{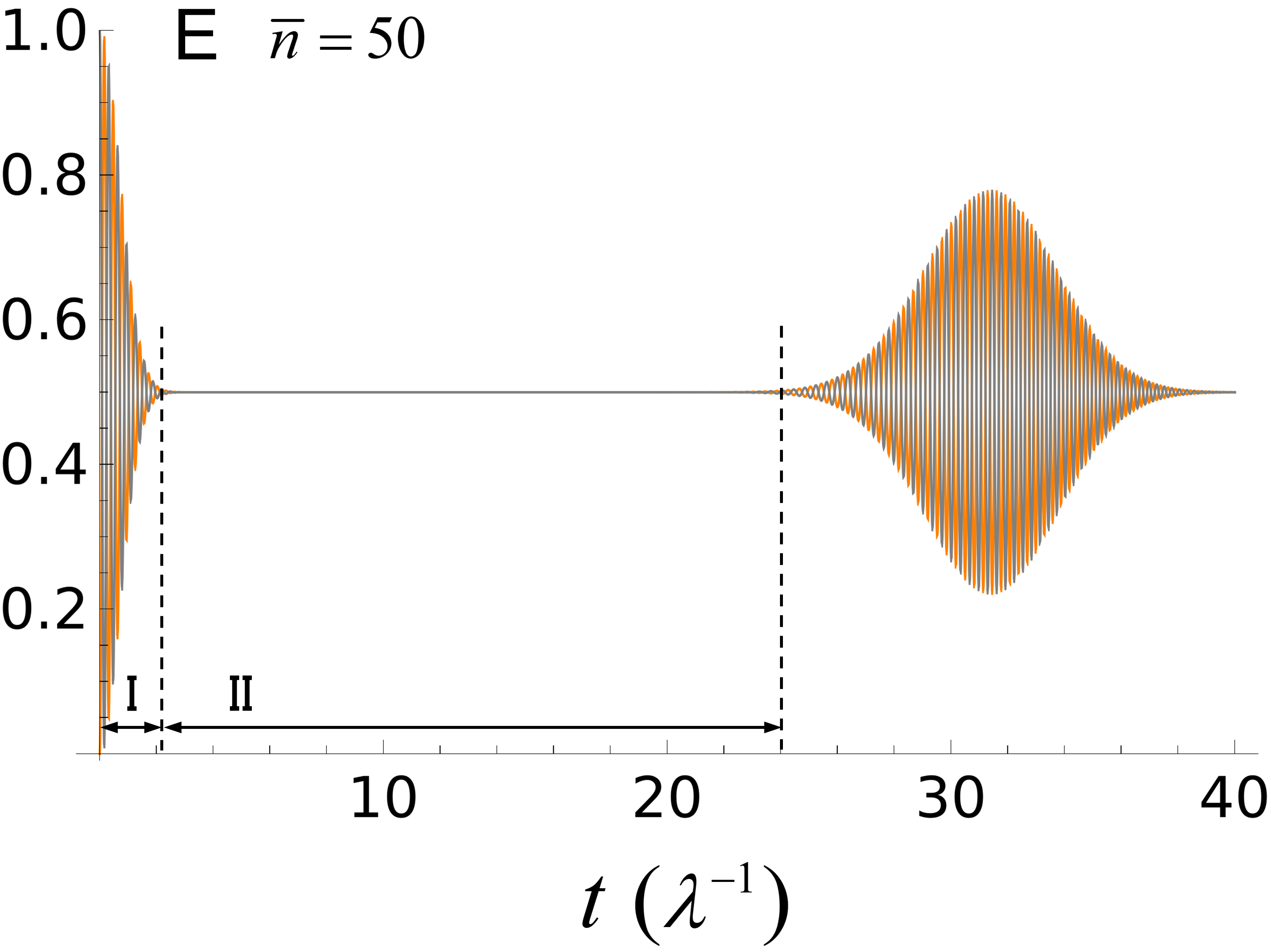}
    \end{minipage}
    \hfill
    \begin{minipage}{0.49\textwidth}
    \centering
    \includegraphics[width=1\textwidth]{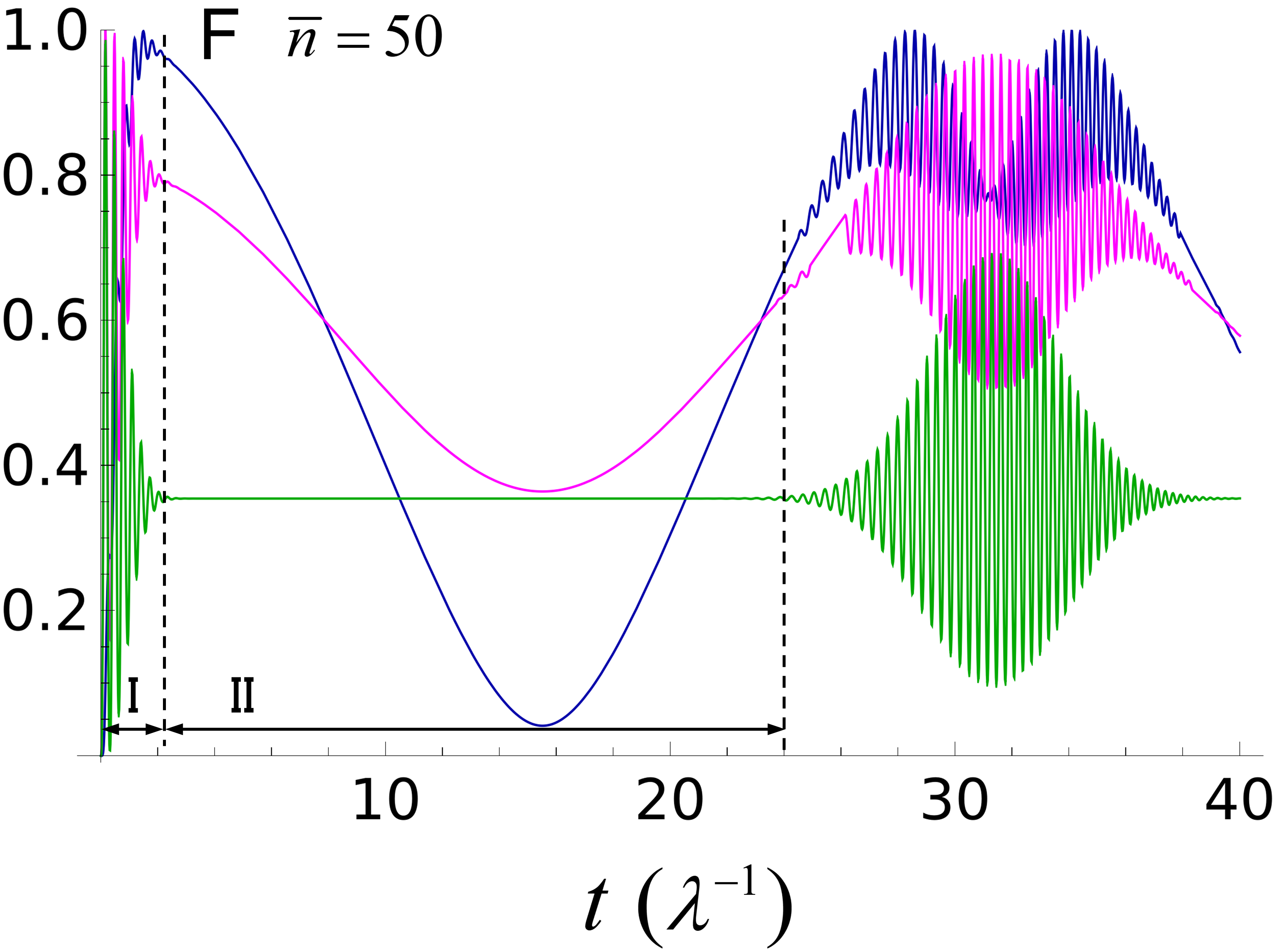}
    \end{minipage}
\caption{Time-dependence, in units of $\lambda^{-1}$, of the atomic probabilities
$P_{\text{atoms}}^{(\alpha)} \left(\ket{\text{sym}},t \right)$ 
and $P_{\text{atoms}}^{(\alpha)} \left(\ket{gg},t \right)$ 
in \eqref{Psymatoms1} and \eqref{Pggatoms1}, respectively, 
and of the entanglements between various components of the total system consisting of the two atoms 
$a$, $b$ and the quantized radiation field $f$ (Section \ref{sec:entanglement} and Figure \ref{fig:entanglement}).
The total system is initially in the state $\ket{g,g,\alpha}$ in \eqref{ic}.
All plots are for $\omega_f=\omega_0=\lambda$, zero detuning ($\Delta  = 0$),  
and for the indicated values of $\bar{n}$.
(A) Probabilities $P_{\text{atoms}}^{(\alpha)} \left(\ket{\text{sym}},t \right)$ 
(orange lines) and 
$P_{\text{atoms}}^{(\alpha)} \left(\ket{gg},t \right)$ (gray lines) of the atomic states 
$\ket{\text{sym}}$ and $\ket{gg}$, respectively. The sum of these probabilities is equal to 1
by normalization.
(B) Entanglements for $\bar{n}=10$:
Entanglement $E(ab-f; t)$ in \eqref{ab-f} between the composite two-atom system $ab$
and the field $f$ (blue lines), entanglement $E(a-bf; t)$ in \eqref{a-bf}
between atom $a$ and the composite (atom $b$)-field system $bf$
(magenta lines), and entanglement of formation $E(a-b; t)$ in \eqref{eof_atoms}
between atoms $a$ and $b$ (green lines).
The colors of the lines in (B) correspond to Figure \ref{fig:entanglement}. 
(C), (D) and (E), (F): same as (A), (B) for $\bar{n}=20$ and $\bar{n}=50$, respectively. 
Parts (E) and (F) highlight intervals in which
$P_{\text{atoms}}^{(\alpha)} \left(\ket{\text{sym}},t \right)$ 
first collapses (I) and then remains constant at $1/2$ (II) before the first revival,
showing the close correlation between the atomic probabilities and the entanglements.}
\label{fig:coherent}  
\end{figure}

\newpage

We now apply these definitions to the total system consisting of the two atoms $a$, $b$ and the 
quantized radiation field $f$. We assume that the total system is in the pure state 
$\ket{ \psi^{(\alpha)}(t) }$ 
in \eqref{evolc} with density operator $\hat{\rho}(t)$ in
\eqref{rho}
and consider the bipartition $ab \hspace{1pt} | \hspace{1pt} f$ of the total system into
the composite two-atom system $ab$ (``atoms") and the field $f$
(see Figure \ref{fig:entanglement}). Using \eqref{vne} the entanglement
$E(ab-f)$ between the two-atom system $ab$ and the field $f$ is given by 
\begin{equation} \label{ab-f}
    E(ab-f) = S_{\text{atoms}} = 
    - \text{tr}\left\{ \hspace{1pt} \hat{\rho}_{\text{atoms}} \log_2 \hat{\rho}_{\text{atoms}} \right\} \, .
\end{equation}
The reduced density operator 
$\hat{\rho}_{\text{atoms}} = \text{tr}_f \left( \hspace{1pt} \hat{\rho} \right)$ from
\eqref{rhoatoms} is represented in the basis $\left\{ \ket{\text{sym}}, \ket{gg} \right\}$ 
by the matrix
\begin{equation} \label{rho_atoms_matrix}
    \rho_{\text{atoms}}(t) =
    \begin{pmatrix}
        P_{\text{atoms}}^{(\alpha)} \left(\ket{\text{sym}},t \right) & \gamma(t) \\[4pt]
        \gamma^*(t) & 1 - P_{\text{atoms}}^{(\alpha)} \left(\ket{\text{sym}},t \right) \hspace{1pt}
    \end{pmatrix}  \, \, ,
\end{equation}
where $P_{\text{atoms}}^{(\alpha)} \left(\ket{\text{sym}},t \right)$ from \eqref{Psymatoms1}
is the probability of finding the two atoms at $t > 0$ in the state $\ket{\text{sym}}$ in \eqref{symatoms}
and
\begin{equation} \label{gamma}
    \gamma(t) := \langle \text{sym} \, | \, \hat{\rho}_{\text{atoms}}(t) \,
    | \, \text{g,g} \rangle \, .
\end{equation}
An explicit expression for $\gamma(t)$ is given in \ref{sec:appC}. 
$E(ab-f)$ in \eqref{ab-f}  
varies between 0 when the atomic and field states are completely disentangled (where $\hat{\rho}_{\text{atoms}}$
describes a pure state) and 1 when the atomic and field states are maximally entangled 
(where both eigenvalues of $\hat{\rho}_{\text{atoms}}$ are equal to $1/2$).
Figure \ref{fig:coherent} shows $E(ab-f)$ in \eqref{ab-f} as a function of 
time $t$ for $\bar{n}=10$, $\bar{n}=20$, and $\bar{n}=50$ 
(blue lines in parts (B), (D), (F), respectively). 

\subsection{Entanglement $E(a-bf)$ between atom $a$ and the composite (atom $b$)-field system $bf$}
\label{sec:a-bf}

Following the definitions in Section \ref{sec:ab-f} we now consider the
bipartition $a \hspace{1pt} | \hspace{1pt} bf$ of the total system into atom $a$ and the 
composite (atom $b$)-field system $bf$ (see Figure \ref{fig:entanglement})
\footnote{Switching atoms $a$ and $b$ does not change the results in this subsection
because the Hamiltonian \eqref{ham} is symmetric under atom exchange.}.
We again assume that the total system is in the pure state $\ket{ \psi^{(\alpha)}(t) }$ in \eqref{evolc}.
Using \eqref{vne} the entanglement between the subsystems $a$ and $bf$ is given by 
\begin{equation} \label{a-bf}
    E(a-bf) = S_a = 
    - \text{tr}\left\{ \hspace{1pt} \hat{\rho}_a \log_2 \hat{\rho}_a \right\} \, ,
\end{equation}
where the reduced density operator $\hat{\rho}_a$ of atom $a$ can be obtained 
as the partial trace of $\hat{\rho}_{\text{atoms}}$ over atom $b$:
\begin{equation} \label{rho_a}
    \hat{\rho}_a = 
\text{tr}_{b,f} \left( \hspace{1pt} \hat{\rho} \right) = \text{tr}_b \left( \hspace{1pt} \hat{\rho}_{\text{atoms}} \right) \, .
\end{equation}
In the basis $\left\{ \ket{g}_a, \ket{r}_a \right\}$ the density operator $\hat{\rho}_a$ 
is given by the matrix
\begin{equation} \label{rho_a_matrix}
    \rho_a(t) =
    \begin{pmatrix}
        1-\frac{1}{2} P_{\text{atoms}}^{(\alpha)} \left(\ket{\text{sym}},t \right) & \frac{1}{\sqrt{2}} \gamma^*(t) \\[4pt]
        \frac{1}{\sqrt{2}} \gamma(t) & \frac{1}{2} P_{\text{atoms}}^{(\alpha)} \left(\ket{\text{sym}},t \right)
    \end{pmatrix} \, \, .
\end{equation}
Figure \ref{fig:coherent} shows $E(a-bf)$ in \eqref{a-bf} as a function of 
time $t$ for $\bar{n}=10$, $\bar{n}=20$, and $\bar{n}=50$ (magenta lines in parts
(B), (D), (F), respectively). 

\subsection{Entanglement of formation $E(a-b)$ between the two atoms $a$ and $b$}
\label{sec:a-b}

Consider again a composite quantum system $Q$ with Hilbert space $\mathbb{H}$
and a bipartition $A | B$ of this system into subsystems $A$ and $B$ 
as discussed at the beginning of Section \ref{sec:ab-f}. 
If the composite system is in a mixed state described by a density operator $\hat{\rho}$
rather than in a pure state $\ket{\psi}$, 
the entanglement can no longer be obtained as the von Neumann entropy of the
subsystems $A$ or $B$. Instead, a quantitative measure for the entanglement 
$E(\hspace{1pt} \hat{\rho})$ is now given by the 
entanglement of formation defined as the minimum average entanglement entropy 
over all pure state decompositions of $\hat{\rho}$, i.e., 
minimized over all ensembles of states $\ket{\psi_i}$ with probabilities $p_i$ 
such that $\hat{\rho} = \sum_i p_i \ketbra{\psi_i}{\psi_i}$
\cite{Bennett1996,Hill1997,Wootters1998,Horodecki2009}.
The entanglement of formation is efficiently computable for two qubits \cite{Wootters1998},
in our case the two two-state atoms $a$ and $b$.
More general multipartite entanglement measures for pure and mixed states are reviewed in
\cite{Horodecki2009}. 

Following reference \cite{Wootters1998} the entanglement of formation
$E(\hspace{1pt} \hat{\rho})$ of an arbitrary 
(possibly mixed) state of two qubits with density operator $\hat{\rho}$ can be calculated 
as follows. Representing $\hat{\rho}$ in the standard basis of a 
two-qubit system by a $4 \times 4$ matrix $\rho$,
one first finds the square roots $\lambda_i \hspace{1pt}$, 
$i=1, \ldots, 4$, 
of the eigenvalues of the $4 \times 4$ matrix
$\rho \tilde{\rho}$ where 
$\tilde{\rho} = (\sigma_y \otimes \sigma_y) \, \rho^{*} (\sigma_y \otimes \sigma_y)$,
$\rho^{*}$ is the complex conjugate of $\rho$ taken in the standard basis,
and $\sigma_y = \begin{pmatrix} 0 & -i \, \\ i & 0 \end{pmatrix}$ is a Pauli matrix
(the tilde indicates a spin flip of the quantum state).
The standard basis is given by 
$\left\{ \ket{gg}, \ket{gr}, \ket{rg}, \ket{rr} \right\}$
in our notation.
Each $\lambda_i$ is a non-negative real number. 
One then finds the concurrence
$C( \hspace{1pt} \rho) = \max
\left\{ 0, \lambda_1 - \lambda_2 - \lambda_3 - \lambda_4 \right\}$
where $\lambda_1$ is the largest of the $\lambda_i \hspace{1pt}$.
The entanglement of formation is then given by \cite{Bennett1996,Hill1997,Wootters1998}
\begin{equation} \label{eof}
    E(\hspace{1pt} \hat{\rho})
    = h \left(\frac{1}{2} + \frac{1}{2}
    \sqrt{1-C(\hspace{1pt} \rho)^2} \right) \, \, ,
\end{equation}
where $h(x) = -x \log_2 x - (1-x) \log_2(1-x)$. 

\begin{figure}
  \centering
  \includegraphics[width=0.68\textwidth]{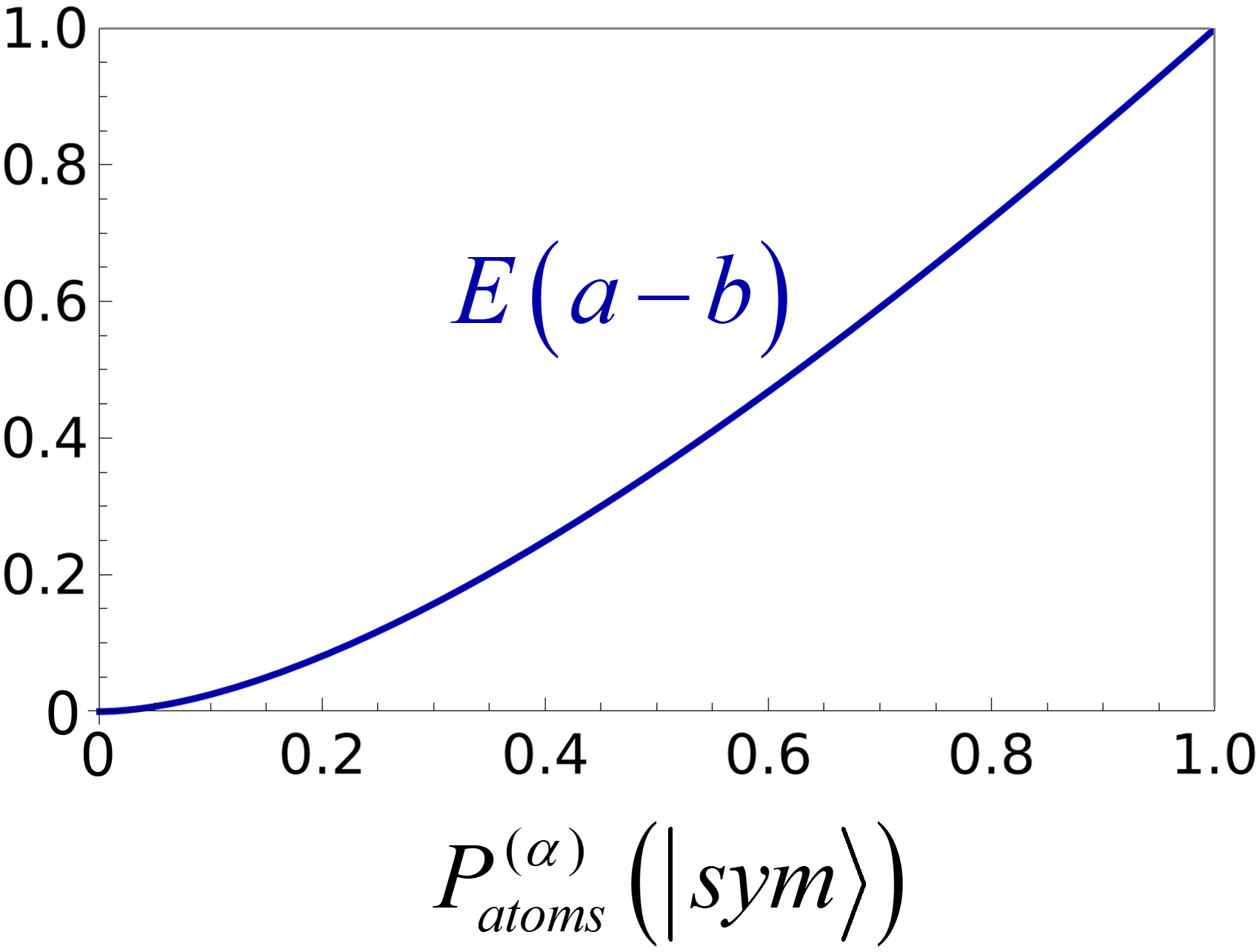}
  \caption{Entanglement of formation $E(a-b)$ between the two atoms $a$ and $b$ 
  in \eqref{eof_atoms} as a function of 
  the probability $P_{\text{atoms}}^{(\alpha)} \left(\ket{\text{sym}} \right)$ 
  of finding the two atoms in the state $\ket{\text{sym}}$ in \eqref{symatoms}.}
  \label{fig:eof}
\end{figure}

We now apply these definitions to the
mixed state of the composite two-atom system $ab$ (``atoms")
represented by the reduced density operator 
$\hat{\rho}_{\text{atoms}}(t) = \text{tr}_f \left\{ \hspace{1pt} \hat{\rho}(t) \right\}$
given in \eqref{rhoatoms}. This results
in the entanglement of formation $E(a-b)$ of 
the two-atom system $ab$, corresponding to the entanglement of atom $a$ with atom $b$
(see Figure \ref{fig:entanglement}).
In the standard two-qubit basis quoted above the density operator
$\hat{\rho}_{\text{atoms}}$ is represented by the matrix 
\begin{equation} \label{rho_atoms_qubit}
    \rho_{\text{atoms}}(t) =
    \begin{pmatrix}
        P_{gg}(t) & \frac{1}{\sqrt{2}} \gamma^*(t) & \frac{1}{\sqrt{2}} \gamma^*(t) & 0 \\[4pt]
        \frac{1}{\sqrt{2}} \gamma(t) & \frac{1}{2} P_{\text{sym}}(t) & \frac{1}{2} P_{\text{sym}}(t) & 0 \\[4pt]
        \frac{1}{\sqrt{2}} \gamma(t) & \frac{1}{2} P_{\text{sym}}(t) & \frac{1}{2} P_{\text{sym}}(t) & 0 \\[4pt]
        0 & 0 & 0 & 0
    \end{pmatrix} \, \, .
\end{equation}
Here, we use the abbreviations $P_{gg}(t) = P_{\text{atoms}}^{(\alpha)} \left(\ket{gg},t \right)$ and
$P_{\text{sym}}(t) = P_{\text{atoms}}^{(\alpha)} \left(\ket{\text{sym}},t \right)$ for ease of notation, 
where $P_{gg}(t) + P_{\text{sym}}(t) = 1$ by normalization.
The rightmost column and the bottom row of the matrix \eqref{rho_atoms_qubit} are filled with zeros 
because the doubly excited state $\ket{rr}$ is not available to the system due to the Rydberg blockade 
between the atoms. 
We find that the matrix $\rho_{\text{atoms}} \cdot \tilde{\rho}_{\text{atoms}}$ has only 1
nonzero eigenvalue given by $P_{\text{sym}}^{\,2}$ resulting in 
$C(\hspace{1pt}\rho_{\text{atoms}}) = P_{\text{sym}}
= P_{\text{atoms}}^{(\alpha)}\left(\ket{\text{sym}} \right)$.
That is, the concurrence is precisely equal to 
the probability of finding the two atoms in the state $\ket{\text{sym}}$ in \eqref{symatoms}.
Using \eqref{eof} we obtain the entanglement of formation as a function of time $t$:
\begin{equation} \label{eof_atoms}
    E(a-b;t)
    = h \left( \frac{1}{2} + \frac{1}{2} 
    \sqrt{1 - \left[ P_{\text{atoms}}^{(\alpha)} \left(\ket{\text{sym}},t \right) \right]^2 } \right) \, \, .
\end{equation}
Clearly, $E(a-b;t)$ is entirely determined by the probability 
$P_{\text{atoms}}^{(\alpha)} \left(\ket{\text{sym}},t \right)$
(Figure \ref{fig:eof}).  
The limiting behaviors are 
$E(a-b) \sim P_{\text{sym}}^{\,2}$ for $P_{\text{sym}} \to 0$
(with a logarithmic correction) and $E(a-b) \to 1$ linearly for 
$P_{\text{sym}} \to 1$.
Figure \ref{fig:coherent} shows $E(a-b;t)$ in \eqref{eof_atoms} as a function of 
time $t$ for $\bar{n}=10$, $\bar{n}=20$, and $\bar{n}=50$ (green lines in parts
(B), (D), (F), respectively).

Figure \ref{fig:coherent} shows that the three forms of entanglement
$E(ab-f,t)$, $E(a-bf,t)$, $E(a-b;t)$ (Figure \ref{fig:entanglement})
and the probability $P_{\text{atoms}}^{(\alpha)} \left(\ket{\text{sym}},t \right)$
are strongly correlated.
The direct correlation between $P_{\text{atoms}}^{(\alpha)} \left(\ket{\text{sym}},t \right)$
and $E(a-b;t)$ is evident by \eqref{eof_atoms} and Figure \ref{fig:eof}.
As expected, all three forms of entanglement vanish for $t=0$ because the total system
is initially prepared in the product state
$\ket{g,g,\alpha}$ in \eqref{ic}. 
Subsequently, $P_{\text{atoms}}^{(\alpha)} \left(\ket{\text{sym}},t \right)$
collapses to a constant value $1/2$ (orange lines in parts (A), (C), (E)). 
Interestingly, $E(ab-f,t)$ and $E(a-bf,t)$ 
(blue and magenta lines in parts (B), (D), (F))
do not remain constant but decrease to relatively low values 
during the period where $P_{\text{atoms}}^{(\alpha)} \left(\ket{\text{sym}},t \right)$
remains constant at $\simeq 1/2$ after the initial collapse. 


\section{Summary and conclusions}
\label{sec:conclusions}
We have extended the Jaynes-Cummings model to study two two-level Rydberg atoms,
with ground states $\ket{g}$ and Rydberg states $\ket{r}$, interacting with 
a quantum single-mode radiation field under perfect Rydberg blockade conditions, 
so that the doubly excited state $\ket{r,r}$ is not accessible to the atoms
(Figures \ref{fig:rb}, \ref{fig:atoms}). 
We find the eigenstates and eigenvalues of the Hamiltonian \eqref{ham}
for general detuning $\Delta = \omega_f - \omega_0$ 
(\eqref{energyasym} - \eqref{evpm})
and $\Delta = 0$ (\eqref{zero32} - \eqref{zero33}),
where $\omega_f$ is the frequency of the field
and $\omega_0$ is the unperturbed transition frequency 
$\ket{g} \leftrightarrow \ket{r}$. 
The eigenstates include the state $\ket{\text{asym}, n} 
= \tfrac{1}{\sqrt{2}} \left(\ket{r,g,n} - \ket{g,r,n} \right)$
in \eqref{asym} for which the atoms are decoupled from the 
radiation field. 

For number (Fock) states of the field, we find explicit expressions for the 
dynamics of the combined system (atoms + field) 
for the initial states $\ket{g,g,n+1}$ (Section \ref{sec:caseB})
and $\ket{r,g,n}$ (Section \ref{sec:caseC}), where $n$ is the number of photons.
For the initial state $\ket{g,g,n+1}$ the system remains in the subspace
spanned by $\ket{\text{sym},n}$ and $\ket{g,g,n+1}$ for all times $t$, 
where $\ket{\text{sym},n} = \tfrac{1}{\sqrt{2}} 
\left(\ket{r,g,n} + \ket{g,r,n} \right)$ defined in \eqref{sym}.
Conversely, for the initial state $\ket{r,g,n}$ the system oscillates between 
$\ket{r,g,n}$ and $\ket{g,r,n}$ in a non-harmonic way 
(see Figure \ref{fig:Cd0} for $\Delta=0$ and Figure \ref{fig:Cd2} for $\Delta \neq 0$),
undergoing collapses and revivals for $\Delta \neq 0$ similarly as for the
Jaynes-Cummings model for one atom \cite{Eberly1980,Larson2021}.
Moreover, 
for the initial state $\ket{g,g,n+1}$ we discuss the entanglement between 
the two-atom system and the quantized field, and describe a scheme
how the extended Jaynes-Cummings model \eqref{ham} can entangle 
atomic systems via photons in principle (see \eqref{Cent} and text below). 

We find explicit expressions for the dynamics of the 
combined system (atoms + field) for the case that the field is in a quantum single-mode 
coherent state (Section \ref{sec:coherent} and Figure \ref{fig:coherent}).
Using these results, we obtain the entanglement between various components of the 
total system consisting of the two Rydberg-interacting atoms $a$, $b$ 
and the quantized radiation field (Figures \ref{fig:entanglement} and \ref{fig:coherent}).
We find distinct correlations between these entanglements and 
the probability $P_{\text{atoms}}^{(\alpha)} \left(\ket{\text{sym}},t \right)$
of finding the atoms in the symmetric state 
$\ket{\text{sym}} = \frac{1}{\sqrt{2}} \left( \ket{r,g} + \ket{g,r} \right)$ 
in \eqref{symatoms}, which are absent
for a system of non-interacting atoms studied by the Tavis-Cummings model (see below). 
The correlations between the atomic states and the atom-field entanglement
is potentially relevant for 
quantum state-transfer between atoms and photons in quantum information processing.

It is interesting to compare the three types of entanglement for the system 
with Rydberg-interacting atoms shown in parts (B), (D), and (F) of Figure \ref{fig:coherent} 
with the corresponding entanglements for a system with non-interacting atoms described by 
the two-atom Tavis-Cummings model (TC). For zero detuning 
and identical coupling constant $\lambda$ between each of the
atoms $a$ and $b$ and the field, the TC model is governed by the Hamiltonian 
\begin{equation} \label{ham_TC}
    \hat H_{TC} = \hbar \omega_f {\hat a}^{\dag} {\hat a} - \frac{1}{2} \hbar \omega_f
    \left( {\hat \sigma}_z^a + {\hat \sigma}_z^b \right)
    \, + \hbar \lambda \left( {\hat \sigma}_{+}^a {\hat a} 
    + {\hat \sigma}_{-}^a {\hat a}^{\dag} 
    + {\hat \sigma}_{+}^b {\hat a} 
    + {\hat \sigma}_{-}^b {\hat a}^{\dag} \right) \, ,
\end{equation}
where $\omega_f$ is again the angular frequency of the quantized radiation field,
which for zero detuning is equal to the atomic transition frequency $\omega_0$. 
The TC Hamiltonian \eqref{ham_TC} is obtained from the Hamiltonian \eqref{ham}
by replacing the projection operators
${\hat P}_g^{\,a} = \ketbra{g_a}{g_a}$ and
${\hat P}_g^{\,b} = \ketbra{g_b}{g_b}$ to the ground states of the atoms 
with the identity operator $\hat{I}$,
so that the doubly excited state $\ket{r,r,\phi_f}$ is accessible in the TC model. 
Entanglements between the atoms and the field in the TC model were studied in 
reference \cite{Tessier2003} in terms of tangles, which is a different 
(albeit closely related, see \ref{sec:appTC})
entanglement measure than the von Neumann entropy and entanglement of formation
discussed in Section \ref{sec:entanglement} and shown in our Figure \ref{fig:coherent}.

\begin{figure}[H]
    \centering
    \begin{minipage}{0.47\textwidth}
    \centering
    \includegraphics[width=1\textwidth]{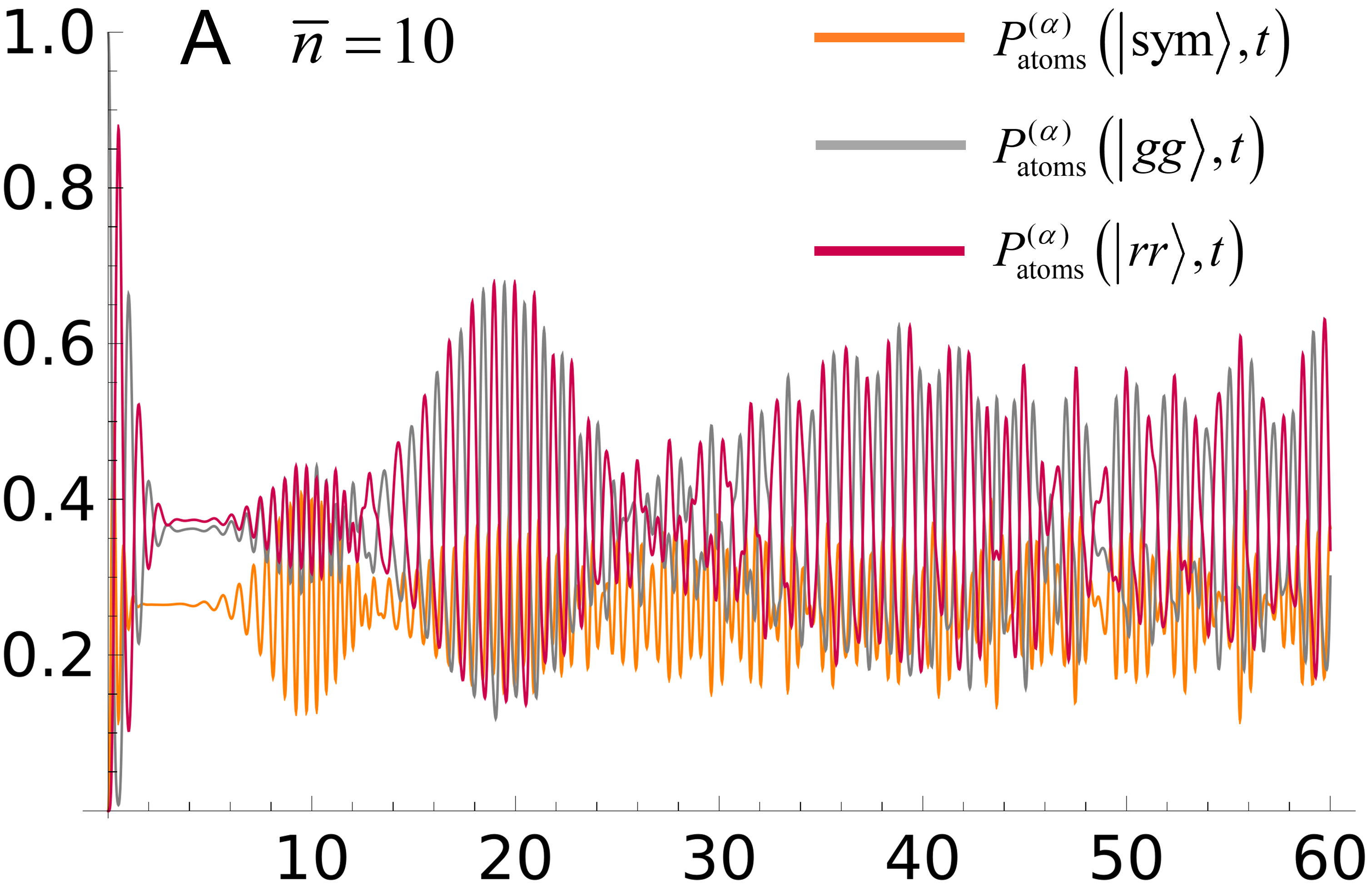}
    \end{minipage}
    \hfill
    \begin{minipage}{0.51\textwidth}
    \centering
    \includegraphics[width=1\textwidth]{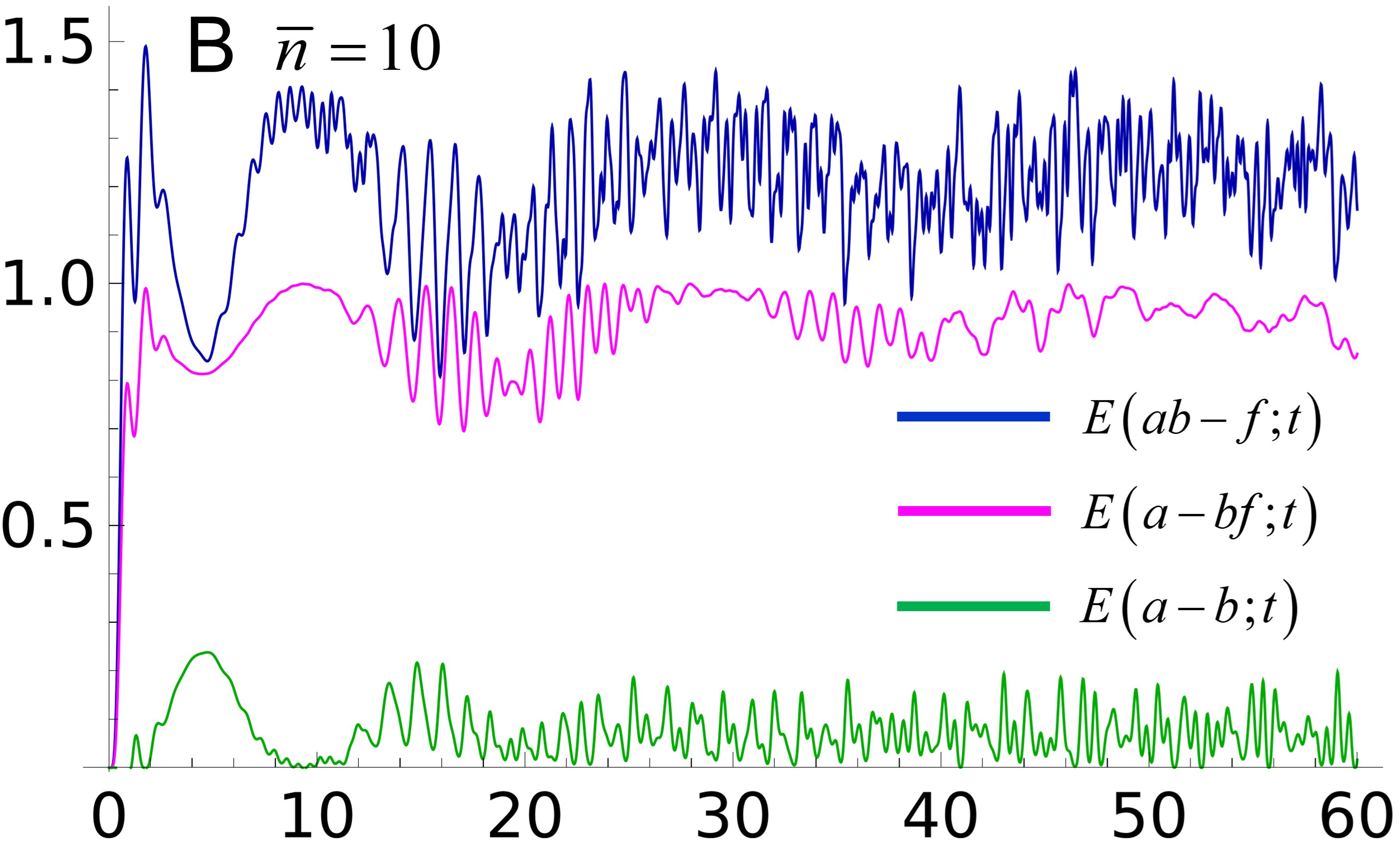}
    \end{minipage}
\par\bigskip
\begin{minipage}{0.47\textwidth}
    \centering
    \includegraphics[width=1\textwidth]{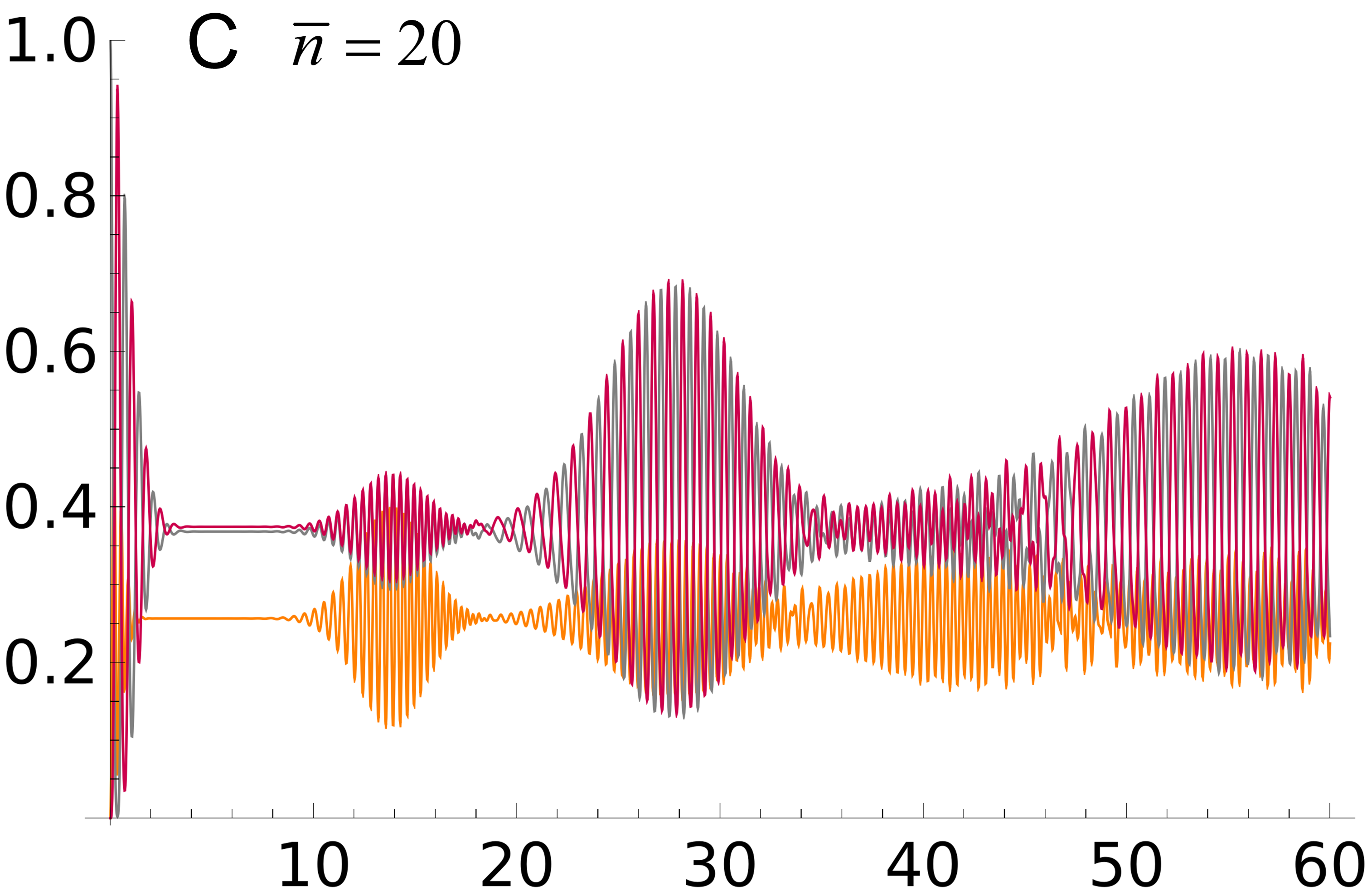}
    \end{minipage}
    \hfill
    \begin{minipage}{0.51\textwidth}
    \centering
    \includegraphics[width=1\textwidth]{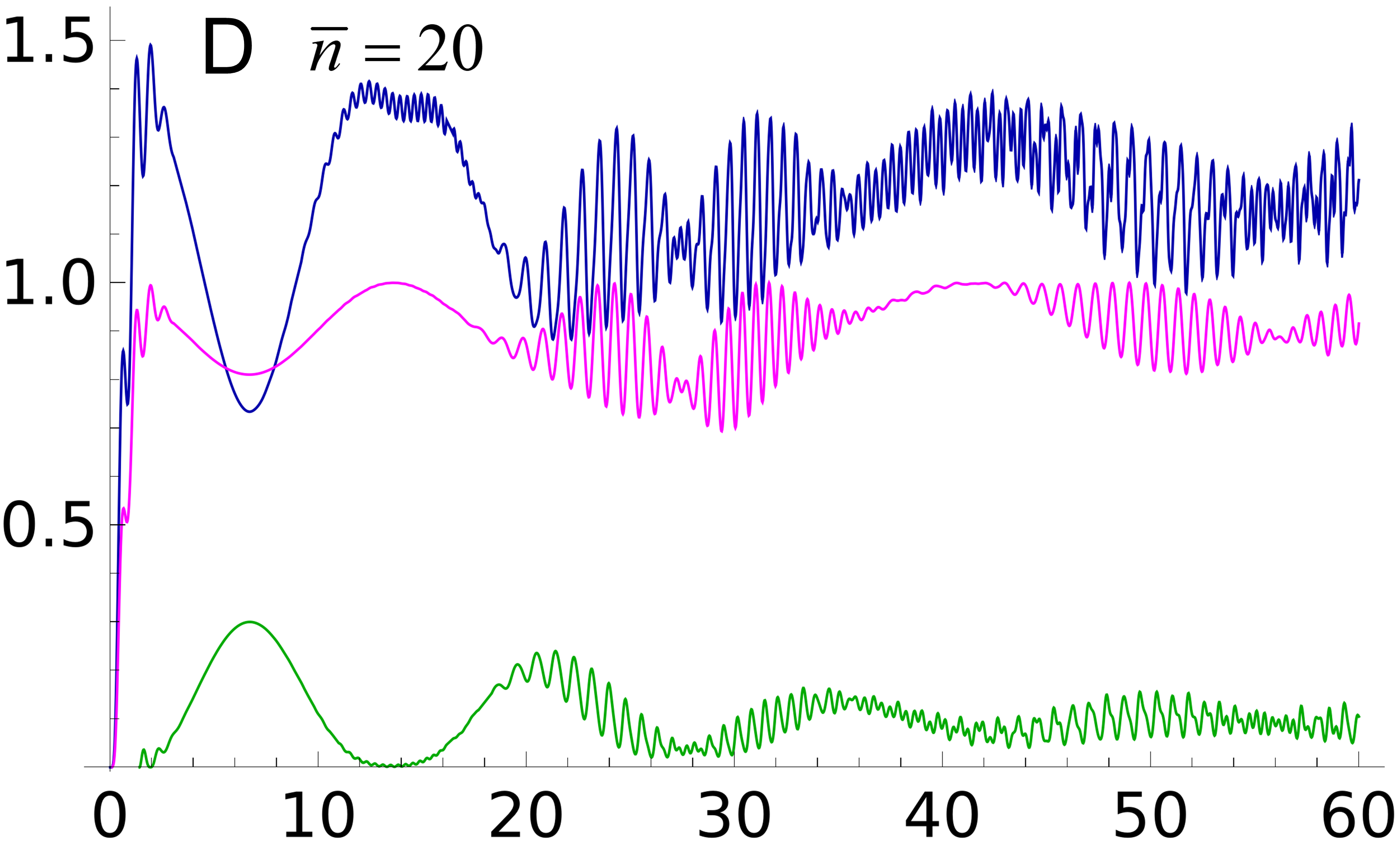}
    \end{minipage}
\par\bigskip
\begin{minipage}{0.47\textwidth}
    \centering
    \includegraphics[width=1\textwidth]{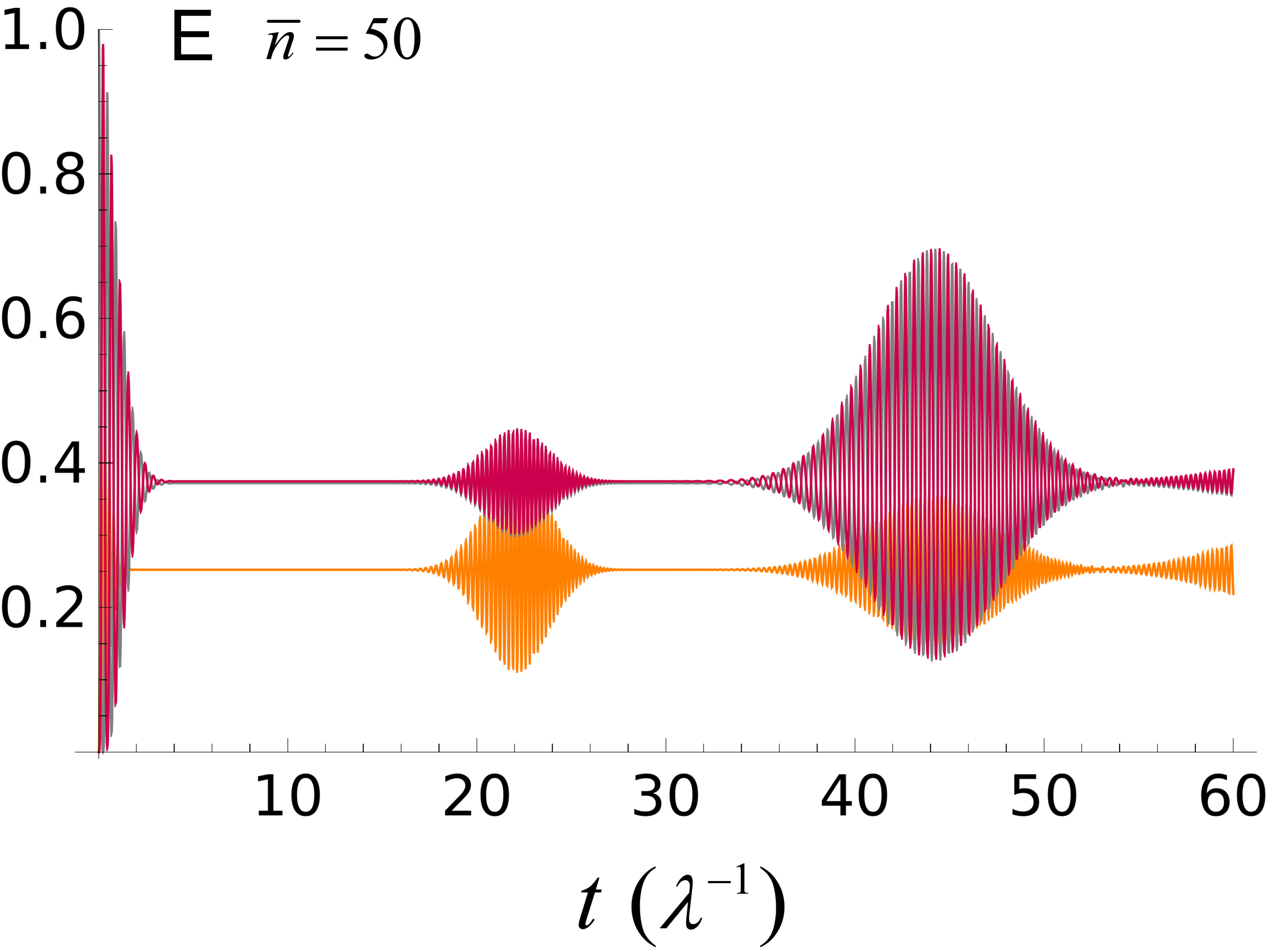}
    \end{minipage}
    \hfill
    \begin{minipage}{0.51\textwidth}
    \centering
    \includegraphics[width=1\textwidth]{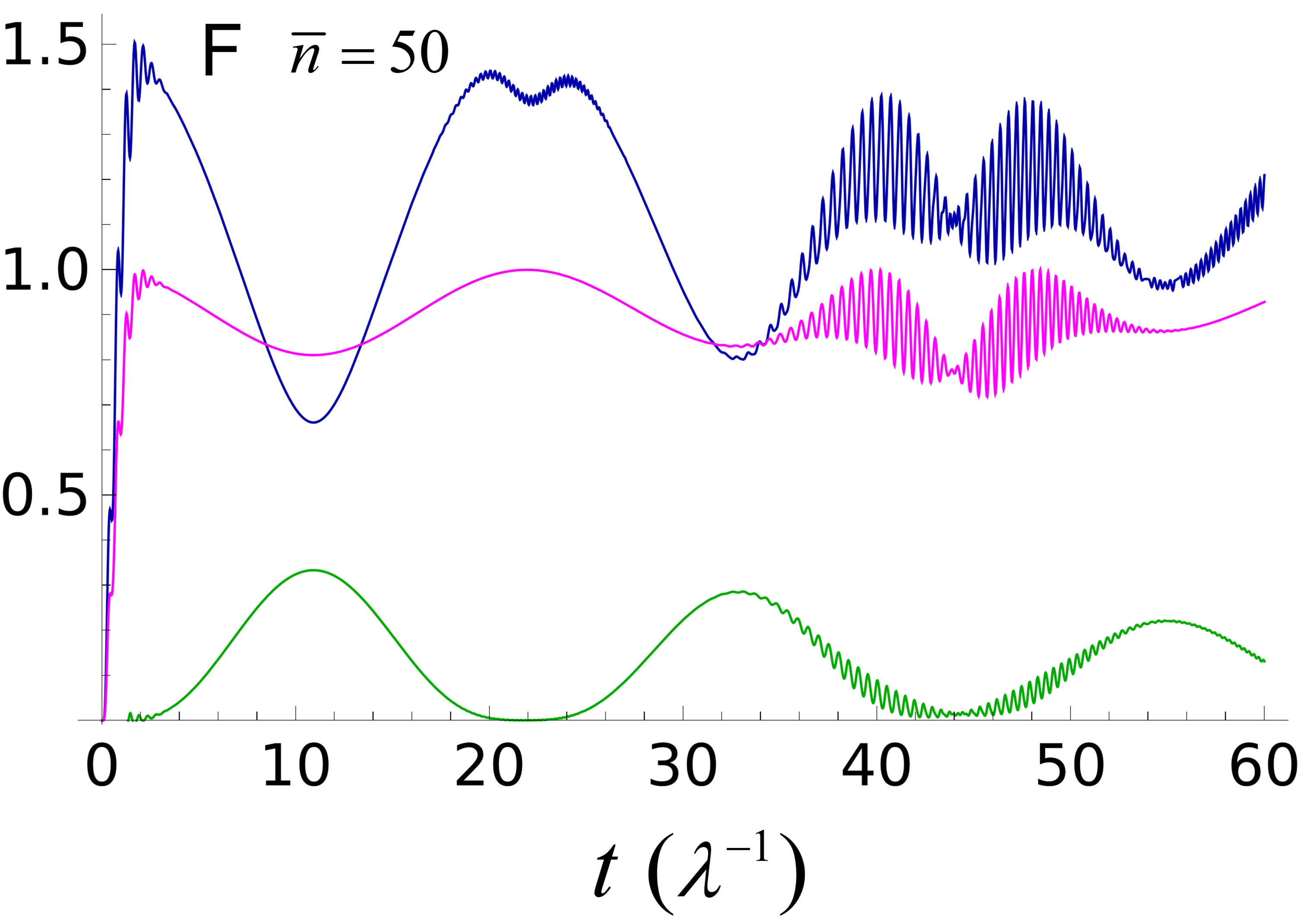}
    \end{minipage}
\caption{Time-dependence, in units of $\lambda^{-1}$, of the atomic probabilities
$P_{\text{atoms}}^{(\alpha)} \left(\ket{\text{sym}},t \right)$,
$P_{\text{atoms}}^{(\alpha)} \left(\ket{gg},t \right)$, and 
$P_{\text{atoms}}^{(\alpha)} \left(\ket{rr},t \right)$
(see \eqref{rho_atoms_matrix_TC} in \ref{sec:appTC}), 
as well as the three types of entanglement $E(ab-f; t)$, $E(a-bf; t)$, and
$E(a-b; t)$ defined in \eqref{ab-f}, \eqref{a-bf}, and \eqref{eof_atoms}, respectively,
obtained for the TC model in \eqref{ham_TC} (compare Figure \ref{fig:coherent}). 
The sum of the atomic probabilities is equal to 1 by normalization.
The parameters, labels and the meaning of the colored lines are the same 
as in Figure \ref{fig:coherent}; the only exception are the probabilities 
$P_{\text{atoms}}^{(\alpha)} \left(\ket{rr},t \right)$ shown as purple lines in 
parts (A), (C), (E), which are absent in Figure \ref{fig:coherent} because the doubly excited 
state $\ket{rr}$ is not accessible for Rydberg-interacting atoms.
As in Figure \ref{fig:coherent}, results are shown for $\bar{n} = 10$, $20$, and $50$
for $\omega_f=\omega_0=\lambda$, zero detuning ($\Delta  = 0$), and for the initial
state $\ket{g,g,\alpha}$. See text and \ref{sec:appTC}.}
\label{fig:coherent_TC}  
\end{figure}

To facilitate a one-to-one comparison of the behavior of the system with
Rydberg-interacting atoms (this work) and independent atoms (TC model), 
Figure \ref{fig:coherent_TC} shows the atomic probabilities  
$P_{\text{atoms}}^{(\alpha)} \left(\ket{\text{sym}},t \right)$,
$P_{\text{atoms}}^{(\alpha)} \left(\ket{gg},t \right)$, and
$P_{\text{atoms}}^{(\alpha)} \left(\ket{rr},t \right)$ 
of the respective atomic states $\ket{\text{sym}}$, $\ket{gg}$, and  $\ket{rr}$ 
(see \eqref{rho_atoms_matrix_TC} in \ref{sec:appTC}), 
as well as the three types of entanglement $E(ab-f; t)$, $E(a-bf; t)$, and
$E(a-b; t)$ defined in \eqref{ab-f}, \eqref{a-bf}, and \eqref{eof_atoms}, respectively,
obtained for the TC model in \eqref{ham_TC}
(see \ref{sec:appTC} for details). 
Figure \ref{fig:coherent_TC} should be 
compared with Figure \ref{fig:coherent}. The parameters, labels,
and the meaning of the colored lines in Figure \ref{fig:coherent_TC} are the same 
as in Figure \ref{fig:coherent}; the only exception are the probabilities 
$P_{\text{atoms}}^{(\alpha)} \left(\ket{rr},t \right)$ shown as purple lines in 
parts (A), (C), (E) in Figure \ref{fig:coherent_TC}, which are absent in 
Figure \ref{fig:coherent} because the doubly excited state $\ket{rr}$
is not accessible for Rydberg-interacting atoms. 

Figures \ref{fig:coherent} and \ref{fig:coherent_TC} show that the time dependence 
of the three types of entanglement in parts (B), (D), and (F)
is notably different for Rydberg-interacting atoms compared to the TC model.
In particular, for the TC model (Figure \ref{fig:coherent_TC}), 
the entanglements are less correlated with the atomic probability
$P_{\text{atoms}}^{(\alpha)} \left(\ket{\text{sym}},t \right)$ 
than for the system with Rydberg-interacting atoms (Figure \ref{fig:coherent}).
Moreover, the atom-atom entanglement $E(a-b; t)$ is significantly
lower for the TC model than for the system with Rydberg-interacting atoms. 
Thus, the Rydberg blockade interaction between the atoms strongly affects entanglement 
in the system overall and increases the entanglement between the atoms.

As mentioned above, in reference \cite{Tessier2003}, entanglements between the atoms and the field 
in the TC model were studied in terms of tangles instead of the von Neumann entropy and entanglement 
of formation shown in Figures \ref{fig:coherent} and \ref{fig:coherent_TC}.
In Figure \ref{fig:tangle} in \ref{sec:appTC} we show tangles 
corresponding to the system in Figure \ref{fig:coherent_TC} 
for $\bar{n}=50$ and $\bar{n}=100$.
The curves for $\bar{n}=50$ in Figure \ref{fig:tangle}A
closely resemble the corresponding curves in Figure \ref{fig:coherent_TC}F, showing 
that our conclusions are independent of the particular entanglement measure.
The curves for $\bar{n}=100$ in Figure \ref{fig:tangle}B exactly reproduce 
the corresponding curves in Fig.~2a in \cite{Tessier2003}. 

Our contribution aims to discuss the basic phenomenon of two Rydberg atoms interacting with a quantum single-mode electromagnetic field under perfect Rydberg blockade conditions, allowing for an essentially analytical treatment. The model may be extended in various ways to deal with more realistic experimental conditions. Firstly, any experimental atom-cavity system is an open system coupled to the environment by loss channels and/or propagating light fields, which can be described by a Lindblad master equation for the density operator $\hat{\rho}$ of the system \cite{Breuer2007}, 
non-Hermitian Hamiltonians \cite {Moiseyev2011}, and input-output theory \cite{Carmichael2008,Kiilerich2019}. 
Secondly, a more realistic model will include 
additional atomic states, e.g., hyperfine ground states of $^{87}$Rb to store qubits \cite{Bluvstein2022,Bluvstein2024}, and multiphoton transitions via intermediate states \cite{Messina2003,Sanchez2016}. 
Furthermore, recent successful experimental realizations of photon-photon quantum gates use electromagnetically induced transparency to map Rydberg interactions to light \cite{Paredes-Barato2014,Tiarks2019}; a realistic description of light-matter interactions of this type in the quantum regime will eventually merge quantum optics 
with a many-body quantum theory of condensed matter physics, 
where the Jaynes-Cummings model as described here may serve as a reference point
\cite{Reiserer2015,Larson2021}.  

\section*{Acknowledgements}
We thank Volker Quetschke for helpful comments and a critical reading of the manuscript.
This work was supported by NSF grants OMA-2231387 and PHY-2012172. 



\appendix

\section{Derivations for Section \ref{sec:caseB}: Case B} \label{sec:appA}

We first derive results for the initial state 
$\ket{ \psi_B^{(n)}(0) }
= \ket{\psi_3^{(n)}}  = \ket{g,g,n+1}$ 
discussed in Section \ref{sec:caseB} (\ref{sec:appsubB}),
and then for the complementary initial state 
$\ket{ \psi_{\beta}^{(n)}(0) } := 
\ket{\text{sym},n} = 
\frac{1}{\sqrt{2}} \left( \ket{ \psi_1^{(n)} } 
+ \ket{ \psi_2^{(n)} } \right)$
(\ref{sec:appsubbeta}). 


\subsection{Initial state \texorpdfstring{$\ket{ \psi_B^{(n)}(0) }
= \ket{\psi_3^{(n)}}  = \ket{g,g,n+1}$}{ }} 
\label{sec:appsubB}

Here we derive the coefficients $\mu_B^{(n)}(t)$, $\nu_B^{(n)}(t)$, $\xi_B^{(n)}(t)$
in \eqref{B1}, \eqref{B3}, \eqref{B4} and the resulting probabilities.  
Using \eqref{comp3} one obtains 
\begin{align}
    \ket{\psi_B^{(n)}(t)} 
    &= \exp \left( - \frac{i}{\hbar} \hat{H} t \right) \ket{\psi_3^{(n)}} \\[2pt] 
    &= \exp \left( - \frac{i}{\hbar} \hat{H} t \right)
    \left[ \sin(\varphi_n) \ket{+,n} + \cos(\varphi_n) \ket{-,n} \right] \\[2pt] 
    &= \sin(\varphi_n) \exp \left( - \frac{i}{\hbar} E_{+}^{(n)} t \right) \ket{+,n}
     + \cos(\varphi_n) \exp \left( - \frac{i}{\hbar} E_{-}^{(n)} t \right) \ket{-,n} \, ,
     \label{psiBapp}
\end{align}
with $\hat{H}$ in \eqref{ham2} and the eigenvalues $E_{\pm}^{(n)}$ in \eqref{energypm}. 
Combining \eqref{psiBapp} with \eqref{psirot3b}, \eqref{psirot3c} one obtains 
\begin{align}
\mu_B^{(n)}(t) &= \left\langle \psi_1^{(n)} \Big| \psi_B^{(n)}(t) \right\rangle \\[2pt] 
& = \sin(\varphi_n) \exp \left( - \frac{i}{\hbar} E_{+}^{(n)} t \right) 
\left\langle \psi_1^{(n)} \Big| +, n \right\rangle
+ \cos(\varphi_n) \exp \left( - \frac{i}{\hbar} E_{-}^{(n)} t \right) 
\left\langle \psi_1^{(n)} \Big| -, n \right\rangle \\[2pt]
& = \frac{1}{\sqrt{2}} \sin(\varphi_n) \cos(\varphi_n)
\left[\exp \left( - \frac{i}{\hbar} E_{+}^{(n)} t \right) - 
\exp \left( - \frac{i}{\hbar} E_{-}^{(n)} t \right) \right] \\[2pt]
& = \frac{1}{\sqrt{2}} \, \frac{1}{2} \sin(2 \varphi_n)
\exp \Big[ -i \hspace{1pt} \Big( \omega_f n + \tfrac{\Delta}{2} \Big) \hspace{1pt} t \Big] 
\left[ \exp \left(-i \Omega_n t\right) 
- \exp \left(i \Omega_n t\right) \right] \\[2pt]
& = \frac{1}{\sqrt{2}} (-i) \sin(2 \varphi_n)
    \exp \Big[ -i \hspace{1pt} \Big( \omega_f n + \tfrac{\Delta}{2} \Big) \hspace{1pt} t \Big] 
    \sin(\Omega_n t) \, , \label{muBapp} \\[10pt]
\nu_B^{(n)}(t) &= \left\langle \psi_2^{(n)} \Big| \psi_B^{(n)}(t) \right\rangle = \mu_B^{(n)}(t) 
\, , \label{nuBapp}
\end{align}
\begin{align}
\xi_B^{(n)}(t) &= \left\langle \psi_3^{(n)} \Big| \psi_B^{(n)}(t) \right\rangle \\[2pt] 
& = \sin(\varphi_n) \exp \left( - \frac{i}{\hbar} E_{+}^{(n)} t \right) 
\left\langle \psi_3^{(n)} \Big| +, n \right\rangle
+ \cos(\varphi_n) \exp \left( - \frac{i}{\hbar} E_{-}^{(n)} t \right) 
\left\langle \psi_3^{(n)} \Big| -, n \right\rangle \\[2pt]
& = \sin^2(\varphi_n) \exp \left( - \frac{i}{\hbar} E_{+}^{(n)} t \right)
+ \cos^2(\varphi_n) \exp \left( - \frac{i}{\hbar} E_{-}^{(n)} t \right) \label{xiBapp} \\[2pt]
& = \exp \Big[ -i \hspace{1pt} \Big( \omega_f n + \tfrac{\Delta}{2} \Big) \hspace{1pt} t \Big] 
\left[\sin^2(\varphi_n) \exp \left(-i \Omega_n t\right) + 
\cos^2(\varphi_n) \exp \left(i \Omega_n t\right) \right] \, .
\end{align}    

\vspace*{2mm}

\noindent
Equations \eqref{muBapp} and \eqref{nuBapp} imply for the corresponding probabilities
\begin{equation}
    P_B^{(n)} \left( \ket{\psi_1^{(n)}}, t \right) 
    = P_B^{(n)} \left( \ket{\psi_2^{(n)}}, t \right)
    = \left| \mu_B^{(n)}(t) \right|^2 
    = \frac{1}{2} \sin^2(2 \varphi_n) \sin^2(\Omega_n t) \, .
\end{equation}
The probability of finding the state
$\ket{\text{sym},n} = \left( \ket{\psi_1^{(n)}} + 
\ket{\psi_2^{(n)}} \right) / \sqrt{2}$ is given by 
\begin{equation} \label{PBsymapp}
    P_B^{(n)}(\text{sym},t) = 2 P_B^{(n)} \left( \ket{ \psi_1^{(n)} }, t \right) 
    = \sin^2(2 \varphi_n) \sin^2(\Omega_n t) \, ,
\end{equation}
as quoted in \eqref{PBsym}.
The probability of finding the complementary state $\ket{\psi_3^{(n)}}$ 
can be calculated similarly, with the result
\begin{align}
    P_B^{(n)} \left( \ket{\psi_3^{(n)}} , t \right)
    &= \left| \xi_B^{(n)}(t) \right|^2 \\[2pt]
    &= 1 - \sin^2(2 \varphi_n) \sin^2(\Omega_n t) \\[2pt]
    &= 1 - P_B^{(n)}(\text{sym},t) \, , \label{PBpsi3app}
\end{align}
as quoted in the text below \eqref{PBsym0}. 


\subsection{Initial state \texorpdfstring{$\ket{ \psi_{\beta}^{(n)}(0) } 
= \ket{\text{sym},n} 
= \frac{1}{\sqrt{2}} \left( \ket{ \psi_1^{(n)} }  
+ \ket{ \psi_2^{(n)} }  \right)$}{ }} 
\label{sec:appsubbeta}

For later use (\ref{sec:appB}), here we quote results for the 
coefficients $\mu_{\beta}^{(n)}(t)$, $\nu_{\beta}^{(n)}(t)$, $\xi_{\beta}^{(n)}(t)$
according to \eqref{evol3} and the resulting probabilities for the initial state 
$\ket{ \psi_{\beta}^{(n)}(0) } 
:= \ket{\text{sym},n}$,
which by \eqref{sigma} and \eqref{PBpsi3app} is complementary 
to the initial state 
$\ket{ \psi_B^{(n)}(0) } = \ket{\psi_3^{(n)}}$
in \ref{sec:appsubB}. Similarly as in \ref{sec:appsubB} one finds 
\begin{align}
    \ket{ \psi_{\beta}^{(n)}(t) } 
    &= \exp \left( - \frac{i}{\hbar} \hat{H} t \right) 
    \frac{1}{\sqrt{2}} \left( \ket{ \psi_1^{(n)} }  
    + \ket{ \psi_2^{(n)} }  \right) \\[2pt]
    &= \exp \left( - \frac{i}{\hbar} \hat{H} t \right)
    \left[ \cos(\varphi_n) \ket{+,n} - \sin(\varphi_n) \ket{-,n} \right] \\[2pt] 
    &= \cos(\varphi_n) \exp \left( - \frac{i}{\hbar} E_{+}^{(n)} t \right) \ket{+,n}
     - \sin(\varphi_n) \exp \left( - \frac{i}{\hbar} E_{-}^{(n)} t \right) \ket{-,n} \, .
     \label{psibetaapp}
\end{align}
Combining \eqref{psibetaapp} with \eqref{psirot3b}, \eqref{psirot3c} one obtains
\begin{align}
\mu_{\beta}^{(n)}(t) &= \left\langle \psi_1^{(n)} \Big| \psi_{\beta}^{(n)}(t) \right\rangle 
= \frac{1}{\sqrt{2}}
\exp \Big[ -i \hspace{1pt} \Big( \omega_f n + \tfrac{\Delta}{2} \Big) \hspace{1pt} t \Big] 
\left[\cos^2(\varphi_n) \exp \left(-i \Omega_n t\right) + 
\sin^2(\varphi_n) \exp \left(i \Omega_n t\right) \right]
\, , \label{mubetaapp1} \\[10pt]
\nu_{\beta}^{(n)}(t) &= 
\left\langle \psi_2^{(n)} \Big| \psi_{\beta}^{(n)}(t) \right\rangle 
= \mu_{\beta}^{(n)}(t) \, , \label{nubetaapp}
\\[10pt]
\xi_{\beta}^{(n)}(t) 
&= \left\langle \psi_3^{(n)} \Big| \psi_{\beta}^{(n)}(t) \right\rangle
= (-i) \sin(2 \varphi_n)
    \exp \Big[ -i \hspace{1pt} \Big( \omega_f n + \tfrac{\Delta}{2} \Big) \hspace{1pt} t \Big] 
    \sin(\Omega_n t) \, . \label{xibetaapp} 
\end{align}    
The probability $P_{\beta}^{(n)}( \text{sym},t)$ 
of finding the system in the state 
$\ket{\text{sym},n} = \frac{1}{\sqrt{2}} \left( \ket{ \psi_1^{(n)} }  
+ \ket{ \psi_2^{(n)} }  \right)$
results as, using
$\ket{ \psi_{\beta}^{(n)}(t) } 
= \sqrt{2} \, \mu_{\beta}^{(n)}(t) \ket{\text{sym},n} + 
\xi_{\beta}^{(n)}(t) \ket{\psi_3^{(n)}}$,
\begin{align} \label{Pbetasymapp}
    P_{\beta}^{(n)}(\text{sym},t) = 2 \left| \mu_{\beta}^{(n)}(t) \right|^2
    &= 1 - \sin^2(2 \varphi_n) \sin^2(\Omega_n t) \\[2pt]
    &= 1 - P_B^{(n)}(\text{sym},t) \, ,
\end{align}
with $P_B^{(n)}(\text{sym},t)$ in \eqref{PBsymapp}.
Thus, the probabilities of finding the state $\ket{\text{sym},n}$ at time $t$
for the complementary initial states 
$\ket{\psi_3^{(n)}}$
and $\ket{\text{sym},n}$ add up to 1 for all $t$.
Furthermore, 
$P_{\beta}^{(n)} \left( \ket{\psi_3^{(n)}} , t \right)
= \left| \xi_{\beta}^{(n)}(t) \right|^2
= 1 - P_{\beta}^{(n)}(\text{sym},t)$,
as expected by normalization. 


\section{Derivations for Section \ref{sec:caseC}: Case C} \label{sec:appB}

For the initial states $\ket{ \psi_C^{(n)}(0) } 
= \ket{ \psi_1^{(n)} } = \ket{r,g,n}$ considered in Section \ref{sec:caseC}
and $\ket{ \psi_2^{(n)} } = \ket{g,r,n}$
the time evolution $\ket{ \psi^{(n)}(t) }$ can be deduced from case A 
(Section \ref{sec:caseA}) and \ref{sec:appsubbeta}, using 
\begin{align}
    \ket{ \psi_1^{(n)} } 
    &= \frac{1}{\sqrt{2}} \left( \ket{ \psi_{\beta}^{(n)}(0) } 
    + \ket{ \psi_A^{(n)}(0) } \right) \label{appC1} \\[4pt]
    \ket{ \psi_2^{(n)} } 
    &= \frac{1}{\sqrt{2}} \left( \ket{ \psi_{\beta}^{(n)}(0) } 
    - \ket{ \psi_A^{(n)}(0) } \right) \, ,
\end{align}
with 
$\ket{ \psi_{\beta}^{(n)}(0) } 
= \frac{1}{\sqrt{2}} \left( \ket{ \psi_1^{(n)} }  
+ \ket{ \psi_2^{(n)} }  \right)$
(\ref{sec:appsubbeta}) and 
$\ket{ \psi_A^{(n)}(0) } 
= \frac{1}{\sqrt{2}} \left( \ket{ \psi_1^{(n)} } - 
\ket{ \psi_2^{(n)} } \right)$
(Section \ref{sec:caseA}).
For the initial state 
$\ket{ \psi_C^{(n)}(0) } = \ket{ \psi_1^{(n)} }$,
\eqref{appC1} implies for the coefficients in \eqref{evol3}
\begin{align} 
    \mu_C^{(n)}(t) &= \frac{1}{\sqrt{2}} \left[ \mu_{\beta}^{(n)}(t) + \mu_A^{(n)}(t) \right] \\[2pt]
    \nu_C^{(n)}(t) &= \frac{1}{\sqrt{2}} \left[ \nu_{\beta}^{(n)}(t) + \nu_A^{(n)}(t) \right] \\[2pt]
    \xi_C^{(n)}(t) &= \frac{1}{\sqrt{2}} \, \xi_{\beta}^{(n)}(t) \, , \, \, 
    \text{using } \xi_A^{(n)}(t)=0 \text{ according to } \eqref{xiA} \, .
\end{align}
Using $\mu_{\beta}^{(n)}(t)$, $\nu_{\beta}^{(n)}(t)$, $\xi_{\beta}^{(n)}(t)$ in
\eqref{mubetaapp1} - \eqref{xibetaapp} and
$\mu_A^{(n)}(t)$, $\nu_A^{(n)}(t)$ in \eqref{munuA}
results in \eqref{C1} - \eqref{C3} quoted in Section \ref{sec:caseC}.  


\section{Derivation of \texorpdfstring{$\gamma(t)$}{ } in \eqref{gamma}}
\label{sec:appC}

\begin{align}
\gamma(t)  &= \langle \text{sym} \, | \, \hat{\rho}_{\text{atoms}}(t) \, | \, g,g \rangle
= \langle \text{sym} \, | \, \text{tr}_f \left\{ \hspace{1pt} \hat{\rho}(t) \right\} | \, g,g \rangle \\[4pt]
&= \sum\limits_{k=0}^{\infty} \left\langle \text{sym}, k \, \Big| \psi^{(\alpha)}(t) \right\rangle
    \left\langle \psi^{(\alpha)}(t) \Big| \, g, g, k \right\rangle \\[2pt]
&= \exp \left( - \left| \alpha \right|^2 \right)  
    \left[ \alpha \, \sigma_B^{(0)}(t) \exp(- i \omega_0 t) 
    + \sum\limits_{k=1}^{\infty} \frac{\alpha^{k+1}}{\sqrt{(k+1)!}} \,
      \sigma_B^{(k)}(t) \, \frac{\alpha^{*k}}{\sqrt{k!}} \, \xi_B^{*(k-1)}(t) \right] \label{appgamma3} \\[4pt]
&= \exp \left( - \left| \alpha \right|^2 \right)  
    \left[ \alpha \, \sigma_B^{(0)}(t) \exp(- i \omega_0 t) 
    + \sum\limits_{k=1}^{\infty} \frac{\left| \alpha \right|^{2k}}{k!}
      \frac{\alpha}{\sqrt{k+1}} \, \sigma_B^{(k)}(t) \, \xi_B^{*(k-1)}(t) \right] 
      \label{appgamma4} \\[2pt]
&= \exp \left( - \left| \alpha \right|^2 \right) \alpha \, (-i) \exp(- i \omega_0 t)
   \sum\limits_{k=0}^{\infty} \frac{\left| \alpha \right|^{2k}}{k!}
      \frac{1}{\sqrt{k+1}} \sin \left(\lambda \sqrt{2(k+1} \, t \right) 
      \cos \left(\lambda \sqrt{2k} \, t \right) \, , \, \, \Delta=0 \, , \label{appgamma5}
\end{align}
where \eqref{appgamma4} holds for general detuning $\Delta$ and \eqref{appgamma5}
for $\Delta=0$.
The first term in \eqref{appgamma3} corresponds to the term $k = 0$ of the sum, using 
\begin{equation}
    \langle k = 0 \, | \, \psi^{(\alpha )}(t) \rangle 
    = \exp \left( - \tfrac{\left| \alpha \right|^2}{2} \right)
    \left[ \exp \left( i \omega_0 t  \right) \ket{g,g} 
    + \alpha \, \sigma_B^{(0)}(t) \ket{\text{sym}} \right]
\end{equation}
with $\ket{ \psi^{(\alpha)}(t) }$ from \eqref{evolc}. 
The result \eqref{appgamma5} is obtained by using the expressions for $\Delta = 0$ of the 
coefficients $\sigma_B^{(n)}(t) = \sqrt{2} \mu_B^{(n)}(t)$ in \eqref{B2}
and $\xi_B^{(n)}(t)$ in \eqref{B5}. 


\section{Tavis-Cummings model} \label{sec:appTC}

Here we briefly review results for the two-atom Tavis-Cummings (TC) model described by the 
Hamiltonian \eqref{ham_TC} as needed to derive the probabilities and entanglements for the TC model 
shown in Figure \ref{fig:coherent_TC}, according to the notation and definitions used 
in Sections \ref{sec:hamiltonian} - \ref{sec:entanglement}.
These results can be deduced from previous work, see, e.g., references \cite{Tessier2003,Zeng2001}. 

The total Hilbert space $\mathbb{H}$ of the TC model \eqref{ham_TC} decouples into 
4-dimensional subspaces
\begin{equation} \label{Hilbert_TC}
\mathbb{H}_n = \text{span} \left\{ \ket{r,r,n-1}, \ket{r,g,n}, \ket{g,r,n}, \ket{g,g,n+1} \right\} \, ,
\, \, n = 1, 2, \ldots
\end{equation}
including the doubly excited atomic state $\ket{r,r,n-1}$
(compare the text in the second paragraph below equation (1)
in \cite{Tessier2003}). 
As in Section \ref{sec:hamiltonian}, 
the index $n$ for $\mathbb{H}_n$ corresponds to the number of photons in the states 
$\ket{r,g,n}$ and $\ket{g,r,n}$ (see text below \eqref{states3}); for $n=0$, the Hilbert space is spanned by
the 3 states $\ket{r,g,0}$, $\ket{g,r,0}$, $\ket{g,g,1}$, and for the index $n=-1$ 
only the state $\ket{g,g,0}$ is available to the system. 
In the basis corresponding to \eqref{Hilbert_TC} 
the Hamiltonian ${\hat H}_{TC}^{\,(n)}$ is represented by a $4 \times 4$ matrix 
\begin{equation} \label{h4matrix_TC}
    {\hat H}_{TC}^{\,(n)} = \hbar \begin{pmatrix}
    n \omega_f & \lambda \sqrt{n} & \lambda \sqrt{n} & 0 \\[4pt]
    \lambda \sqrt{n} & n \omega_f & 0 & \lambda \sqrt {n+1} \\[4pt]
    \lambda \sqrt{n} & 0 & n \omega_f & \lambda \sqrt {n+1} \\[4pt]
    0 & \lambda \sqrt {n+1} & \lambda \sqrt {n+1} &  n \omega_f \end{pmatrix} \, \, , \, n = 1, 2, \ldots
\end{equation}
For the initial state $\ket{\psi^{(n)}_{TC}(0)} = \ket{g,g,n+1}$ at $t=0$ (corresponding to case B 
discussed in Section \ref{sec:caseB}) the time evolution of the system is given by
\begin{align} \label{evol_TC}
    \ket{\psi^{(n)}_{TC}(t)} & = 
    \exp\left(-\frac{i}{\hbar} \hat{H}_{TC\,} t \right) \ket{g,g,n+1} \nonumber \\[1mm]
    & = \rho_n(t) \ket{r,r,n-1} + \sigma_n(t) \ket{ \text{sym},n }
    + \xi_n(t) \ket{g,g,n+1} \, \, , \, n = 1, 2, \ldots
\end{align}
where
\begin{align}
    \rho_n(t) &= \exp (-i \omega_f n t) \, a(n) \, b(n) \left( \cos\left(\Omega_n^{TC} t\right) -1 \right) \, , \\[6pt] 
    \sigma_n(t) &= -i \exp (-i \omega_f n t) \, a(n) \sin\left(\Omega_n^{TC} t\right) \, , \\[6pt]
    \xi_n(t) &= \exp (-i \omega_f n t) \left( a(n)^2 \cos\left(\Omega_n^{TC} t\right) + b(n)^2 \right) \, ,  
\end{align} 
with $a(n)=\sqrt{(n+1)/(2n+1)}$, $b(n) = \sqrt{n/(2n+1)}$, and the Rabi frequency 
of the Tavis-Cummings model
\footnote{Note that $\Omega_n^{TC}$ in \eqref{Rabi_TC} is different from the Rabi frequency 
$\Omega_n(0)$ in \eqref{zero31} by an extra factor of 2 in front of $n$.}
\begin{equation} \label{Rabi_TC}
    \Omega_n^{TC} = \lambda \sqrt{2 (2n+1)} \, \, .
\end{equation}
The normalization $\left| \hspace{1pt} \rho_n(t) \right|^2 
+ \left| \hspace{1pt} \sigma_n(t) \right|^2 + \left| \hspace{1pt} \xi_n(t) \right|^2 = 1$
is fulfilled due to $a(n)^2+b(n)^2=1$.

If the two atoms are initially in their ground states $\ket{g}$ 
and the field is in a quantum single-mode coherent state $\ket{\alpha}$ as in \eqref{cs},
the initial state 
$\ket{\psi_{TC}^{(\alpha)}(0)} = \ket{g,g,\alpha}$ evolves to
\begin{equation} \label{evolc_TC} 
    \ket{ \psi_{TC}^{(\alpha)}(t) } =  
    \exp \left( - \frac{\left| \alpha \right|^2}{2} \right)  
    \left[ \exp(i \omega_0 t) \ket{g,g,0} +
    \sum\limits_{n=0}^{\infty} \frac{\alpha^{n+1}}{\sqrt{(n+1)!}} 
    \ket{\psi^{(n)}_{TC}(t)} \right] \, , \, \, t>0 \, ,
\end{equation}
with $\ket{\psi^{(n)}_{TC}(t)}$ in \eqref{evol_TC} (compare \eqref{evolc}). 
The reduced density operator for the atoms, $\hat{\rho}^{TC}_{\text{atoms}}(t)$,
is given by the partial trace of the density operator $\hat{\rho}_{TC}(t)$
of the total system (atoms + field) in the pure state 
$\ket{ \psi_{TC}^{(\alpha)}(t) }$ 
over the field $\ket{\phi_f}$, i.e., 
\begin{equation} \label{rhoatoms_TC}
    \hat{\rho}^{TC}_{\text{atoms}}(t) = \text{tr}_f \left\{ \hspace{1pt}\hat{\rho}_{TC}(t) \right\}
    = \sum\limits_{k=0}^{\infty} \langle k \, | \, \psi_{TC}^{(\alpha)}(t) \rangle
    \langle \psi_{TC}^{(\alpha)}(t) \, | \, k \rangle \, ,
\end{equation}
where $\ket{k}$, $k=0,1,2,\ldots$ (compare \eqref{rho} and \eqref{rhoatoms}). 
In the basis $\left\{ \ket{rr}, \ket{\text{sym}}, \ket{gg} \right\}$,
with $\ket{\text{sym}} = \frac{1}{\sqrt{2}} \left( \ket{r,g} + \ket{g,r} \right)$,
the reduced density operator for the atoms is represented by the matrix
\begin{equation} \label{rho_atoms_matrix_TC}
    \rho^{TC}_{\text{atoms}}(t) =
    \begin{pmatrix}
        P_{\text{atoms}}^{(\alpha)} \left(\ket{rr},t \right) & \theta^*(t) & \phi(t) \\[4pt]
        \theta(t) & P_{\text{atoms}}^{(\alpha)} \left(\ket{\text{sym}},t \right) & \gamma(t) \\[4pt]
        \phi^*(t) & \gamma^*(t) & P_{\text{atoms}}^{(\alpha)} \left(\ket{gg},t \right)
        \hspace{1pt}
    \end{pmatrix}  \, \, .
\end{equation}
The diagonal elements in \eqref{rho_atoms_matrix_TC} are the probabilities 
of the respective two-atom states 
$\ket{rr}$, $\ket{\text{sym}}$, and $\ket{gg}$, 
and the off-diagonal elements are the coherences, e.g., 
$\gamma(t) = \langle \text{sym} \, | \, \hat{\rho}^{TC}_{\text{atoms}}(t) \,
    | \, \text{gg} \rangle$. 
The coherences $\theta(t)$, $\phi(t)$ are related to the doubly excited state $\ket{rr}$.
Equation \eqref{rho_atoms_matrix_TC} should be compared with the 
reduced density operator $\rho_{\text{atoms}}(t)$ in 
\eqref{rho_atoms_matrix}, in which the state $\ket{rr}$ is absent.
We note that the analytical expressions of $\gamma(t)$ and of the probabilities 
$P_{\text{atoms}}^{(\alpha)} \left(\ket{\text{sym}},t \right)$,
$P_{\text{atoms}}^{(\alpha)} \left(\ket{\text{gg}},t \right)$
in \eqref{rho_atoms_matrix_TC} are different from the ones in
\eqref{rho_atoms_matrix}.
In the standard two-qubit basis
$\left\{ \ket{gg}, \ket{gr}, \ket{rg}, \ket{rr} \right\}$ 
the density operator
$\hat{\rho}^{TC}_{\text{atoms}}$ is represented by the matrix (compare \eqref{rho_atoms_qubit})
\begin{equation} \label{rho_atoms_qubit_TC}
    \overline{\rho}^{TC}_{\text{atoms}}(t) =
    \begin{pmatrix}
        P_{gg}(t) & \frac{1}{\sqrt{2}} \gamma^*(t) & \frac{1}{\sqrt{2}} \gamma^*(t) & \phi^*(t) \\[4pt]
        \frac{1}{\sqrt{2}} \gamma(t) & \frac{1}{2} P_{\text{sym}}(t) & \frac{1}{2} P_{\text{sym}}(t) & 
        \frac{1}{\sqrt{2}} \theta(t) \\[4pt]
        \frac{1}{\sqrt{2}} \gamma(t) & \frac{1}{2} P_{\text{sym}}(t) & \frac{1}{2} P_{\text{sym}}(t) & 
        \frac{1}{\sqrt{2}} \theta(t) \\[4pt]
        \phi(t) & \frac{1}{\sqrt{2}} \theta^*(t) & \frac{1}{\sqrt{2}} \theta^*(t) &  P_{rr}(t)
        \end{pmatrix} \, \, ,
\end{equation}
where the overbar indicates that the matrix is with respect to the standard basis.
We use the same type of abbreviation for the probabilities as quoted below \eqref{rho_atoms_qubit}
with $P_{gg}(t) + P_{\text{sym}}(t) + P_{rr}(t) = 1$ by normalization.
The three types of entanglement $E(ab-f; t)$, $E(a-bf; t)$, and $E(a-b; t)$ shown in 
Figure \ref{fig:coherent_TC} for the TC model were calculated using the density matrices 
$\rho^{TC}_{\text{atoms}}(t)$ and $\overline{\rho}^{TC}_{\text{atoms}}(t)$ in 
\eqref{rho_atoms_matrix_TC} and \eqref{rho_atoms_qubit_TC}, respectively, 
as described in Section \ref{sec:entanglement}.

\begin{figure}[H]
  \centering
  \includegraphics[width=0.80\textwidth]{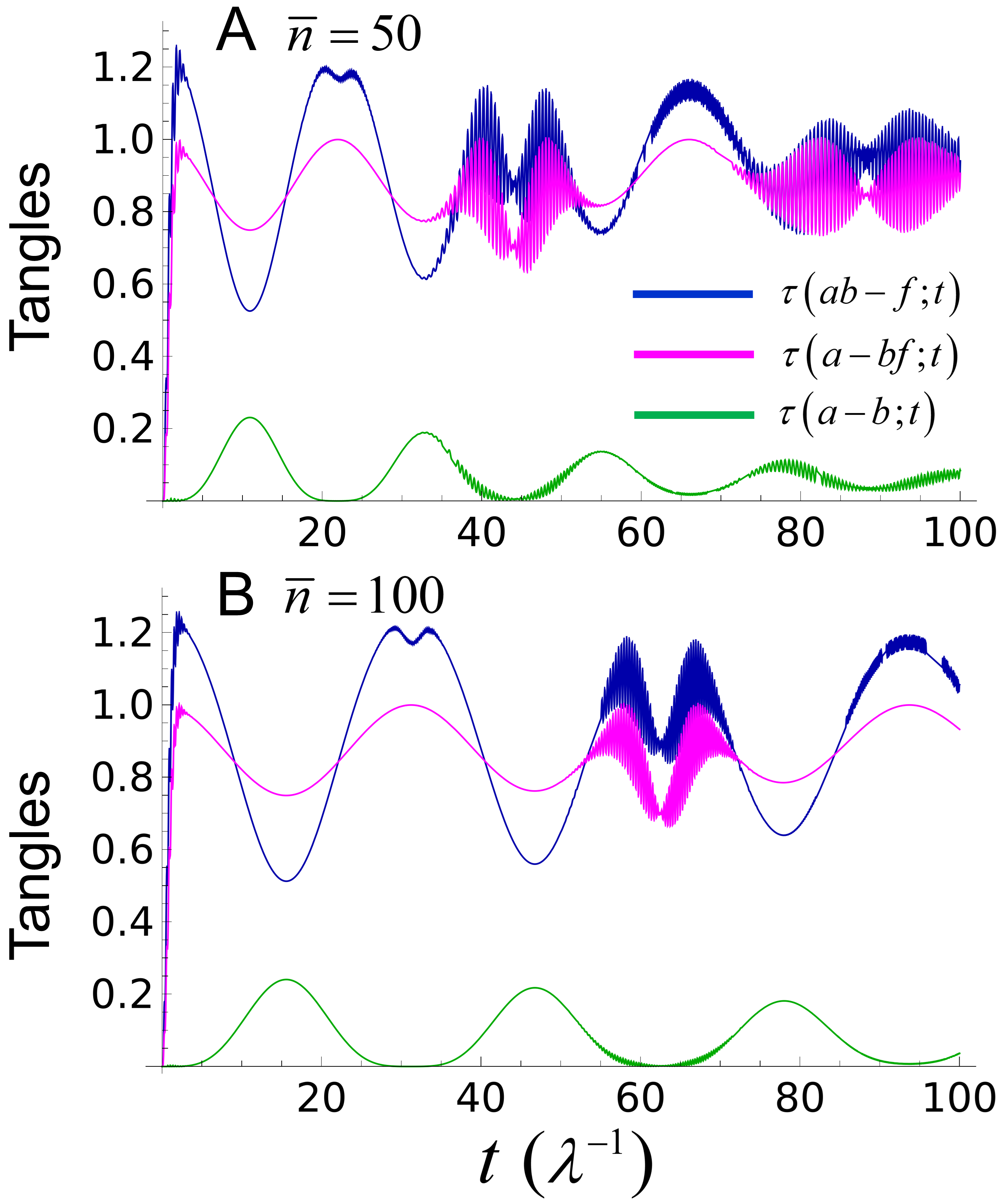}
  \caption{Tangles in \eqref{tau_ab-f} - \eqref{tau_a-b} for (A) $\bar{n}=50$ and (B) $\bar{n}=100$
  for the same system as considered in Figure \ref{fig:coherent_TC}.} 
  \label{fig:tangle}
\end{figure}

In reference \cite{Tessier2003}, entanglements between the atoms and the field in the TC model 
were studied in terms of tangles, which is a different 
entanglement measure than the von Neumann entropy and entanglement of formation
discussed in Section \ref{sec:entanglement} and shown in our Figures \ref{fig:coherent}
and \ref{fig:coherent_TC} (see equations (7), (8), and (2) in \cite{Tessier2003}). 
In terms of tangles, the entanglement between the two-atom system $ab$ and the 
field $f$ is given by (instead of $E(ab-f; t)$ in \eqref{ab-f})
\begin{equation} \label{tau_ab-f}
    \tau(ab-f;t) = 
    2 \left[ 1 - \text{tr} \left\{ \left(\overline{\rho}^{TC}_{\text{atoms}}(t)\right)^2 \right\} \right] \, .
\end{equation}
The entanglement between atom $a$ and the (atom $b$)-field system $bf$ is given 
by (instead of $E(a-bf; t)$ in \eqref{a-bf})
\begin{equation} \label{tau_a-bf}
    \tau(a-bf;t) =
    2 \left[ 1 - \text{tr} \left\{ \left(\rho^{TC}_a(t)\right)^2 \right\} \right] \, ,
\end{equation}
where the reduced density matrix $\rho^{TC}_a$ of atom $a$ is the 
partial trace of $\overline{\rho}^{TC}_{\text{atoms}}(t)$ over atom $b$
(compare \eqref{rho_a}).
Finally, the atom-atom entanglement is given by the tangle
(instead of $E(a-b; t)$ in \eqref{eof_atoms})
\begin{equation} \label{tau_a-b}
    \tau(a-b;t) = C\left( \hspace{1pt} \overline{\rho}^{TC}_{\text{atoms}}(t) \right)^2 \, ,
\end{equation}
where 
$C\left( \hspace{1pt} \overline{\rho}^{TC}_{\text{atoms}}(t) \right)
= \max \left\{ 0, \lambda_1 - \lambda_2 - \lambda_3 - \lambda_4 \right\}$
is the concurrence for the density matrix $\overline{\rho}^{TC}_{\text{atoms}}(t)$
(see Section \ref{sec:a-b}; here, 3 of the 4 $\lambda_i$'s are nonzero).
For reference purposes, Figure \ref{fig:tangle} shows the tangles in 
\eqref{tau_ab-f} - \eqref{tau_a-b} for $\bar{n}=50$ and $\bar{n}=100$
for the same system as considered in Figure \ref{fig:coherent_TC}. 
The curves for $\bar{n}=50$ in Figure \ref{fig:tangle}A
closely resemble the corresponding curves in Figure \ref{fig:coherent_TC}F, and
the curves for $\bar{n}=100$ in Figure \ref{fig:tangle}B exactly reproduce 
the corresponsing curves in Fig.~2a in \cite{Tessier2003}. 

\bibliographystyle{elsarticle-num} 
\bibliography{references}






\end{document}